\documentclass[apj]{emulateapj}

\newcommand{\ADD}[1]{
\ifmmode #1 
\else #1 
\fi
}
\newcommand{\DEL}[1]{} 

\newcommand{\Nco}{\ensuremath{N_\mathrm{CO}}}
\newcommand{\Tco}{\ensuremath{T_\mathrm{CO}}}
\newcommand{\vturb}{\ensuremath{v_\mathrm{turb}}}
\newcommand{\Tbg}{\ensuremath{T_\mathrm{BG}}}
\newcommand{\Nh}{\ensuremath{N_\mathrm{H}}}
\newcommand{\Nhco}{\ensuremath{N_\mathrm{H,4.67}}}
\newcommand{\Nhx}{\ensuremath{N_\mathrm{H,X}}}
\newcommand{\Nhsi}{\ensuremath{N_\mathrm{H,9.7}}}
\newcommand{\LIR}{\ensuremath{L_\mathrm{IR}}}

\newcommand{\Lco}{\ensuremath{L_\mathrm{CO}}}
\newcommand{\AKARI}{\textit{AKARI}}
\newcommand{\Spitzer}{\textit{Spitzer}}
\newcommand{\WISE}{\textit{WISE}}
\newcommand{\HII}{H\,\textsc{ii}}
\newcommand{\psqcm}{\ensuremath{\mathrm{cm^{-2}}}}
\newcommand{\kmps}{\ensuremath{\mathrm{km~s^{-1}}}}

\received{}
\revised{}
\accepted{}

\shorttitle{NIR CO Absorption in AGN Tori}
\shortauthors{Baba et al.}

\begin{document}

\title{The Near-Infrared CO Absorption Band as a Probe to the Innermost Part of an AGN Obscuring Material}


\author{Shunsuke Baba}
\affiliation{Institute of Space and Astronautical Science, Japan Aerospace Exploration Agency, 3-1-1 Yoshinodai, Chuo-ku, Sagamihara, Kanagawa 252-5210, Japan}
\affiliation{Department of Physics, Graduate School of Science, The University of Tokyo, 7-3-1 Hongo, Bunkyo-ku, Tokyo 113-0033, Japan}

\author{Takao Nakagawa}
\affiliation{Institute of Space and Astronautical Science, Japan Aerospace Exploration Agency, 3-1-1 Yoshinodai, Chuo-ku, Sagamihara, Kanagawa 252-5210, Japan}

\author{Naoki Isobe}
\affiliation{Institute of Space and Astronautical Science, Japan Aerospace Exploration Agency, 3-1-1 Yoshinodai, Chuo-ku, Sagamihara, Kanagawa 252-5210, Japan}

\author{Mai Shirahata}
\affiliation{Institute of Space and Astronautical Science, Japan Aerospace Exploration Agency, 3-1-1 Yoshinodai, Chuo-ku, Sagamihara, Kanagawa 252-5210, Japan}

\begin{abstract}
We performed a systematic analysis of the 4.67~\micron\ CO ro-vibrational absorption band toward nearby active galactic nuclei (AGNs) and analyzed the absorption profiles of ten nearby galaxies collected from the \AKARI\ and \Spitzer\ spectroscopic observations that show the CO absorption feature by fitting a plane-parallel local thermal equilibrium gas model.
We found that CO gas is warm (200--500~K) and has a large column density ($\Nh\gtrsim10^{23}~\psqcm$).
The heating of the gas is not explicable by either UV heating or shock heating because these processes cannot represent the large column densities of the warm gas.
Instead, X-ray photons from the nuclei, which can produce large columns of warm gas with up to $\Nh\sim10^{24}~\psqcm$, are the most convincing power source.
The hydrogen column density estimated from the CO band is smaller than that inferred from X-ray observations.
These results indicate that the region probed by the near-infrared CO absorption is in the vicinity of the nuclei and is located outside the X-ray emitting region.
Furthermore, the covering factors nearly unity required by the observed deep absorption profiles suggest that the probed region is close to the continuum source, which can be designated as the inner rim of the obscuring material around the AGN.
\end{abstract}


\section{INTRODUCTION}

Active galactic nuclei (AGNs) show a wide diversity of observational characteristics in their spectra.
AGN spectra differ primarily in terms of optical broad emission lines, the presence or absence of which are used to classify the AGNs as types 1 or 2, respectively.
This dichotomy has been attributed to a viewing angle effect caused by a putative AGN torus, an optically and geometrically thick torus-shaped dusty cloud that obscures direct emission from the nuclear region when the AGN is viewed edge-on \citep[AGN unified model,][]{Antonucci_1993}.
To understand the characteristics of AGNs, it is important to observe AGN tori and verify the AGN unified model.
However, because of their small sizes on parsec scales, it is difficult to directly image AGN tori.
Recent millimeter to sub-millimeter interferometric observations with the Atacama Large Millimeter Array of carbon monoxide (CO) pure rotational emission lines revealed the presence of gas concentrated near central nuclei \citep[e.g.,][]{Garcia-BurilloEtAl_2016}, but the highest spatial resolution that can be achieved with such observations is about several parsecs, even in the nearest AGNs.
Thus, in distant galaxies millimeter and sub-millimeter emission lines are not suitable for resolving AGN tori from their hosts.
An alternative observing method that can be applied to large numbers of AGNs is required.

The strategy we employ in this paper is based on spectroscopy of the CO fundamental ro-vibrational absorption band centered at 4.67~\micron\ ($v=1\leftarrow0$, $\Delta J=\pm1$).
Using the bright near-IR radiation from the central region as the background continuum, this technique can observe foreground molecular gas clouds with an effectively high spatial resolution at the parsec scale because of to the compactness of the near-IR emitting region.
Furthermore, because this band contains multiple lines with different rotation levels in a narrow wavelength range, it is possible to obtain information on the gas excitation state from one observation.
In this respect, the near-IR CO absorption band is preferable to the (sub-)millimeter CO pure rotational emission lines, which are easily affected by contamination from the host galaxy and cannot be observed simultaneously.

\defcitealias{ShirahataEtAl_2013}{S13}
\defcitealias{SpoonEtAl_2004b}{S04}
\citet{GeballeEtAl_2006} and \citet[hereafter \citetalias{ShirahataEtAl_2013}]{ShirahataEtAl_2013} observed the absorption band toward the heavily obscured ultra-luminous infrared galaxy (ULIRG) IRAS 08572$+$3915 using the United Kingdom 3.8-m Infrared Telescope (UKIRT) and the 8.2-m Subaru telescope, respectively.
Strong absorption lines were detected up to high rotational levels ($J\le17$), with the resulting population diagram showing the presence of large columns of warm molecular gas in the line of sight.
\citet[hereafter \citetalias{SpoonEtAl_2004b}]{SpoonEtAl_2004b} observed another obscured ULIRG IRAS 00182$-$7112 using the \textit{Spitzer Space Telescope} and also detected strong CO absorption.
Although their observation did not resolve the multiple lines owing to insufficient spectral resolution, they analyzed the entire absorption profile using a plane-parallel local thermal equilibrium (LTE) gas model \citep{Cami_2002} and also found that the gas is warm and has a large column density.
Based on the high temperatures and large column densities, both authors argued that the observed gas should be in the vicinity of the dominant nuclear power source.
\citetalias{ShirahataEtAl_2013} also proposed that the warm gas is heated by X-ray radiation from an AGN engine.

This CO absorption feature, however, does not always appear in all type-2 AGNs, in which the putative torus should be seen edge-on.
\citet{LutzEtAl_2004} observed nearby 19 Seyfert 1 (Sy1) and 12 Seyfert 2 (Sy2) galaxies using \textit{Infrared Space Observatory} (\textit{ISO}), but none showed the CO feature.
\citet{LahuisEtAl_2007} detected similar warm molecular gas toward obscured (U)LIRGs through the mid-IR absorption bands of $\mathrm{C_2H_2}$, HCN, and $\mathrm{CO_2}$ but concluded that the gas is unlikely to be associated with the material surrounding AGNs because these molecules would be rapidly destroyed in an intense X-ray field.
The two above studies controvert the hypothesis that CO absorption probes warm gas near the central region.

To assess the location of the region probed by CO ro-vibrational absorption, in this study we analyzed space telescope observations of the CO feature toward ten nearby AGNs and compared the results with the results from other X-ray and mid-IR observations.
Such a systematic analysis of the CO absorption profile had not previously been performed, although detection of the feature has been reported in some objects \citep{ImanishiEtAl_2008c, ImanishiEtAl_2010c, SpoonEtAl_2005}.
Together with a description of observations and data reduction, the selection of our targets is explained in Section \ref{sec:obs}.
The method used to analyze the CO absorption profile is described in Section \ref{sec:ana}, followed by presentation of results in Section \ref{sec:res}.
We discuss these results and compare them with other observations in Section \ref{sec:dis} and, finally, we give our conclusion in Section \ref{sec:con}.

\section{TARGETS, OBSERVATIONS, AND DATA REDUCTION}\label{sec:obs}

We used spectroscopic observations carried out with the \AKARI\ satellite \citep{MurakamiEtAl_2007} and the \textit{Spitzer Space Telescope} \citep{WernerEtAl_2004} to collect targets that show CO absorption.
\AKARI\ and \Spitzer\ have near- and mid-IR spectrometers, respectively, which cover complementary redshift ranges.
Because we were not able to obtain information on the longward continuum level over the CO absorption from the \AKARI\ observations themselves, \Spitzer\ data were used to complement the spectrum in longer wavelengths.
For these sources, we scaled the two spectra using \WISE\ catalog magnitudes as reference points.
In the following, we describe in detail the observations and data reduction techniques and present the spectra of the targets.

\subsection{AKARI}

We searched AGNs showing CO absorption from the archival data of the \AKARI\ mission program AGNUL (P.I. T. Nakagawa).
The program conducted many spectroscopic observations of nearby AGNs and ULIRGs using the Infrared Camera (IRC) in the NG grism mode.
Almost all of the observations were performed through a $1\arcmin\times1\arcmin$ aperture, and thus constituted slitless spectroscopy.
The NG grism mode covers a wavelength range from 2.5 to 5.0~\micron.
Although its spectral resolution in general depends on the spatial extent of the target, if it is a point source, the resolution can be given as $R=33.3\lambda$, where $\lambda$ is the observed wavelength in \micron\ \citep{OnakaEtAl_2007, OhyamaEtAl_2007e}.
The redshift range within which it is possible to observe the band center of the CO feature is $z<0.07$.
The AGNUL program also carried out observations in another dispersion mode (NP), but we excluded these from our sample because the spectral resolution of that mode was insufficient for the following analysis.
The observation period of the program is divided into two parts: a cryogenic phase and a post-cryogenic phase.
In this study, we scoped only cryogenic observations, which had been calibrated better than the post-cryogenic observations.
A study using post-cryogenic data will be presented in a forthcoming paper.

Under the above conditions, 50 ULIRGs were observed, eight of which are within $z<0.07$.
We found that the six ULIRGs listed in the upper part of Table \ref{tab:targets} show CO absorption.
The other two ULIRGs are Mrk 231 and IRAS 05189$-$2524, which are classified as Sy1 and Sy2, respectively \citep{VeilleuxEtAl_1995}.
Table \ref{tab:targets} also presents the redshift, optical classification, and IR and X-ray AGN signatures of the six CO ULIRGs.
Although some of these are not classified as Seyferts, either IR or X-ray diagnostics suggest that they are AGN hosts, and we therefore assumed that all six ULIRGs harbor an AGN and used them as targets for analysis.
IRAS 23128$-$5919 is a merging system with a nuclear separation of 5\arcsec\ \citep{DucEtAl_1997}.
The southern nucleus of the galaxy was detected in hard X-rays with \textit{Chandra} and believed to be an obscured AGN based on the observed X-ray hardness ratio \citep{IwasawaEtAl_2011}.

\begin{deluxetable*}{cccccccl} 
\tablecaption{Basic target data\label{tab:targets}}
\tablewidth{0pt}
\tablehead{
\colhead{Group} & \colhead{Object} & \colhead{$z$} & 
\colhead{$\log\LIR$} & \colhead{Optical} &
\multicolumn{2}{c}{AGN Sign} & 
\colhead{Ref.} \\
\cline{6-7} 
\colhead{} & \colhead{} & \colhead{} & 
\colhead{($L_\sun$)} & \colhead{Class} & 
\colhead{IR} & \colhead{X-ray} & \colhead{}}
\startdata
\AKARI\   & IRAS 06035$-$7102 & 0.0797 & 12.2       & LI          & \checkmark  & \nodata        & 1;    6, 7; --- \\
          & IRAS 08572$+$3915 & 0.0583 & 12.1       & LI          & \checkmark  & \checkmark     & 2, 3; 7, 8; 9   \\
          & UGC 5101          & 0.0392 & 12.0       & LI          & \checkmark  & \checkmark     & 3;    7, 8; 9   \\
          & Mrk 273           & 0.0373 & 12.2       & Sy2         & \checkmark  & \checkmark     & 2, 3; 6, 7; 9   \\
          & IRAS 19254$-$7245 & 0.0616 & 12.1       & Sy2         & \checkmark  & \checkmark     & 1, 4; 7, 8; 10  \\
          & IRAS 23128$-$5919 & 0.0448 & 12.0       & \HII/Sy2/LI & ---         & \checkmark     & 1;    6;    9   \\
\hline
\Spitzer\ & IRAS 00182$-$7112 & 0.3270 & 12.9       & LI          & \checkmark  & \checkmark     & 5;    8;    11  \\
          & IRAS 00397$-$1312 & 0.2617 & 13.0       & \HII        & \checkmark  & ---            & 2;    8;    12  \\
          & IRAS 00406$-$3127 & 0.3424 & 12.8       & Sy2         & \checkmark  & \nodata        & 4;    7;    --- \\
          & IRAS 13352$+$6402 & 0.2366 & 12.5       & ?           & \checkmark  & \nodata        & ---;  7;    --- 
\enddata
\tablecomments{Column 1: target group.
Column 2: object name.
Column 3: redshift taken from the PSCz catalog \citep{SaundersEtAl_2000}.
Column 4: logarithm of the infrared (8--1000~\micron) luminosity in units of the solar luminosity $L_\sun$ derived from \citet{SandersEtAl_1996}: $\LIR=2.1\times10^{39}\times D_\mathrm{L}^2\times(13.48f_{12}+5.16f_{25}+2.58f_{60}+f_{100})~\mathrm{erg~s^{-1}}$, where $D_\mathrm{L}$ is the distance in Mpc, and $f_{12}$, $f_{25}$, $f_{60}$, and $f_{100}$ are \textit{IRAS} fluxes in Jy. In calculating $D_\mathrm{L}$, $H_0=70~\kmps~\mathrm{Mpc}^{-1}$, $\Omega_\mathrm{m}=0.3$, and $\Omega_\Lambda=0.7$ are adopted. The \textit{IRAS} fluxes are taken from \citet{SandersEtAl_2003q}, \citet{KimEtAl_1998}, or the \textit{IRAS} Faint Source Catalog Version 2.0. For objects having upper limits in the \textit{IRAS} fluxes, we evaluated the upper and lower limits of \LIR\ by assuming an actual flux equal to the upper limit and a zero value, respectively. Those upper and lower limits are quite close, with a difference less than 0.16~dex. Ultimately, the average of the two limits was adopted.
Column 5: optical spectral classification. ``LI'', ``Sy2'', ``\HII'', and ``?'' denote LINER, Seyfert 2, \HII\ region, and no optical classification, respectively.
Columns 6 and 7: IR and X-ray AGN signatures, respectively. Check: present. Dash: absent. Dots: no data. The IR signature is based on the low equivalent widths of the 3.3 and/or 6.2~\micron\ PAH emissions. The X-ray signature is based on a hard photon index and/or strong iron K lines.
Column 8: references for columns 5, 6, and 7 with semicolons as delimiters.
1: \citet{DucEtAl_1997}.
2: \citet{VeilleuxEtAl_1999a}.
3: \citet{VeilleuxEtAl_1995}.
4: \citet{AllenEtAl_1991}.
5: \citet{ArmusEtAl_1989}.
6: \citet{ImanishiEtAl_2010c}.
7: \citet{SargsyanEtAl_2011}.
8: \citet{ImanishiEtAl_2008c}.
9: \citet{IwasawaEtAl_2011}.
10: \citet{BraitoEtAl_2009}.
11: \citet{NandraEtAl_2007}.
12: \citet{NardiniEtAl_2011}. 
}
\end{deluxetable*}

Table \ref{tab:akari} summarizes observational information obtained from \AKARI/IRC.
In the NG mode, a spectrum is contaminated by second-order light at wavelengths longer than 4.9~\micron.
A correction for this effect was established by \citet{BabaEtAl_2016}, whom we followed in performing data reduction.
One-dimensional raw spectra were extracted using the official IRC Spectroscopy Toolkit Version 20150331 in the standard manner.
To minimize uncertainty in the wavelength calibration, the wavelength origin was adjusted from the \ADD{value reported by the toolkit} by a few pixels based on the positions of features such as the 3.3~\micron\ PAH emission band, H\,{\footnotesize I} Br$\alpha$ and Br$\beta$ emission lines, and 4.26~\micron\ CO$_2$ absorption band.
The wavelength dependence of the refractive index of the grism material was included in the wavelength calibration.
Contamination from the second-order light was correctly removed in the flux calibration.
Because the southern and northern parts of IRAS 23128$-$5919 were barely resolved in the two-dimensional spectral image, we selectively extracted flux from the southern part, which is known to be an AGN host as mentioned above.
IRAS 06035$-$7102, IRAS 08572$+$3915, Mrk 273, and IRAS 19254$-$7245 also have double disks or nuclei, which we were not able to resolve because they either have small separations or are aligned in the dispersion direction of the two-dimensional spectral images.

\begin{deluxetable}{ccc}
\tablecaption{\AKARI/IRC observation log\label{tab:akari}}
\tablewidth{0pt}
\tablehead{
\colhead{Object} & \colhead{Observation ID} & \colhead{Observation Date}
}
\startdata
IRAS 06035$-$7102 & 1100130.1 & 2007 Mar 11 \\
IRAS 08572$+$3915 & 1100049.1 & 2006 Oct 26 \\
UGC 5101          & 1100134.1 & 2007 Apr 22 \\
Mrk 273           & 1100273.1 & 2007 Jun  8 \\
IRAS 19254$-$7245 & 1100132.1 & 2007 Mar 30 \\
IRAS 23128$-$5919 & 1100294.1 & 2007 May 10 
\enddata
\end{deluxetable}

\subsection{Spitzer}

\citet{SpoonEtAl_2005} reported that four ULIRGs observed with the Infrared Spectrometer (IRS) onboard \Spitzer\ \citep{HouckEtAl_2004} show CO absorption.
The lower part of Table \ref{tab:targets} presents basic information on these four ULIRGs.
These ULIRGs are, except for IRAS 00406$-$3127, not optically classified as Seyferts but show IR AGN signatures.
Thus, we investigated these ULIRGs in addition to the \AKARI\ targets.
The \Spitzer\ targets are systematically more luminous than the \AKARI\ targets\DEL{, reflecting the selection bias that more luminous objects are more famous objects and reported earlier with higher priority}.
Table \ref{tab:spitzer} summarizes the \Spitzer/IRS observational information and tabulates the observations used to complement the spectra of the \AKARI\ targets.

The spectra around the CO absorption were obtained in the IRS SL2 mode, which covers wavelengths from 5.21 to 7.56~\micron\ with a spectral resolution $R=16.5\lambda$ at a slit width of 3\farcs6 \citep{IRS_2011}.
This wavelength range corresponds to the redshift range $z=0.12$--0.62.
The spectral resolution is lower than that of the \AKARI/IRC NG mode but sufficient for the following analysis.
The calibrated \Spitzer\ spectra of the targets were downloaded from the IRS Enhanced Products on the \Spitzer\ Heritage Archive.

\begin{deluxetable}{ccc}
\tablecaption{\Spitzer/IRS observation log\label{tab:spitzer}}
\tablewidth{0pt}
\tablehead{
\colhead{Object} & \colhead{AOR Key} & \colhead{Observation Date}
}
\startdata
IRAS 06035$-$7102 W  &  4969728 & 2004 Apr 14 \\
IRAS 08572$+$3915 NW &  4972032 & 2004 Apr 15 \\
UGC 5101             &  4973056 & 2004 Mar 23 \\
Mrk 273              &  4980224 & 2004 Apr 14 \\
IRAS 19254$-$7245 S  & 12256512 & 2005 May 30 \\
IRAS 23128$-$5919 S  &  4991744 & 2004 May 11 \\
\hline
IRAS 00182$-$7112    &  7556352 & 2003 Nov 14 \\
IRAS 00397$-$1312    &  4963584 & 2004 Jan  4 \\
IRAS 00406$-$3127    & 12258816 & 2005 Jul 11 \\
IRAS 13352$+$6402    & 12258560 & 2005 Mar 20 
\enddata
\end{deluxetable}

\subsection{Scaling \ADD{to} the \WISE\ Photometry}

\ADD{
Because the \AKARI/IRC spectra of the \AKARI\ targets lacked the longward part of the CO absorption owing to redshift, we supplemented them with \Spitzer/IRS ones to estimate the longward continuum levels.
To reduce the effect of the different aperture sizes of \AKARI\ and \Spitzer, we scaled the fluxes of the two spectra so that they match with the \WISE\ photometry \citep{WrightEtAl_2010a}.
The procedure was as follows.
Profile-fit magnitudes were taken from the AllWISE catalog and converted into fluxes in Jy based on the zero magnitudes presented by \citet{JarrettEtAl_2011a}, with color corrections taken into account.
The color correction factor for the $W1$ band was calculated by integrating the product of the $W1$ spectral response function and the relative shape of the \AKARI\ spectrum, and those factors for the $W3$ and $W4$ bands were similarly obtained using the relative shape of the \Spitzer\ spectrum.
The $W2$ band flux cannot put clear constraint because the band protrudes from the \AKARI\ wavelength coverage.
We thus evaluated only the lower limit for the factor, which resulted in the flux upper limit.
The \AKARI\ spectrum was scaled so that its flux density at the isophotal wavelength of the $W1$ band ($\lambda_\mathrm{obs}=3.35~\micron$) hits the color-corrected $W1$ flux, and confirmed not to pass over the $W2$ upper limit.
The \Spitzer\ spectrum was scaled so that its flux densities at the $W3$ and $W4$ isophotal wavelengths ($\lambda_\mathrm{obs}=11.56$, 22.09~\micron) fit the corrected fluxes of the two bands.
}

\begin{figure*}
  \centering
  \begin{tabular}{@{}c@{}c@{}c@{}}
    \includegraphics[width=0.33\hsize]{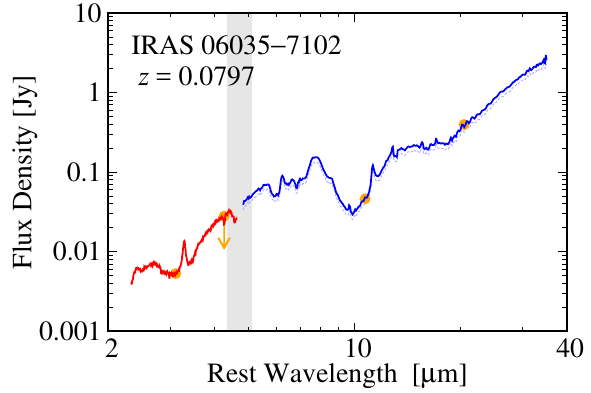}   &
    \includegraphics[width=0.33\hsize]{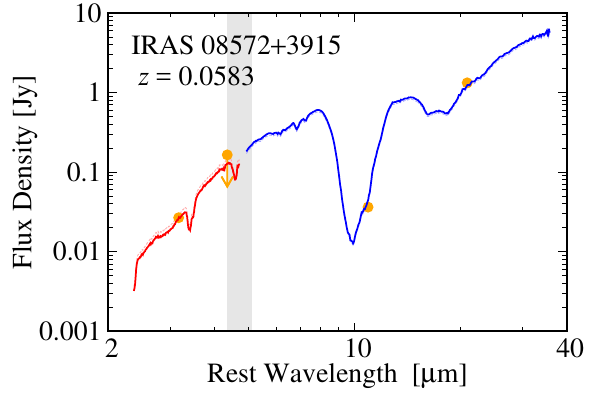}   &
    \includegraphics[width=0.33\hsize]{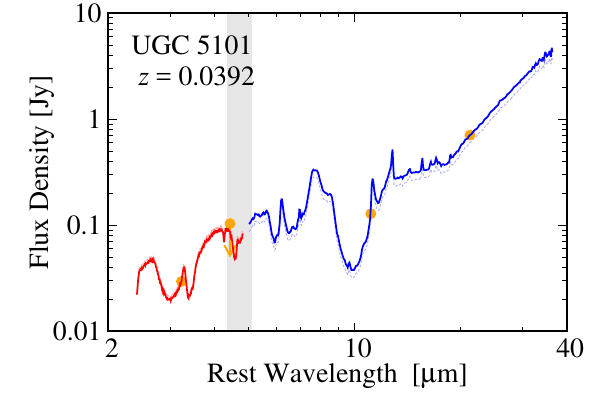}          \\
    \includegraphics[width=0.33\hsize]{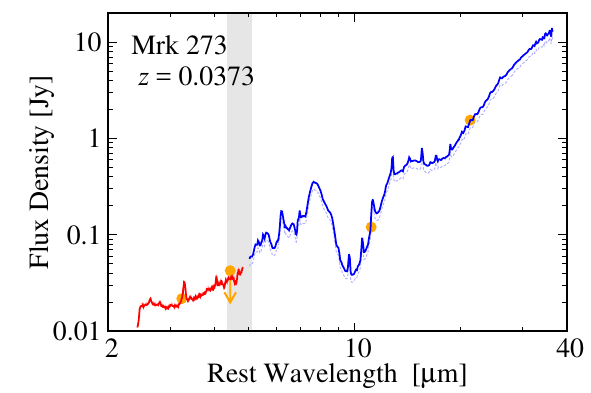}           &
    \includegraphics[width=0.33\hsize]{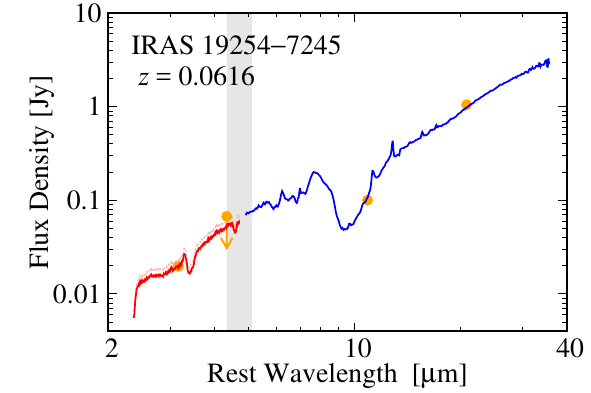}   &
    \includegraphics[width=0.33\hsize]{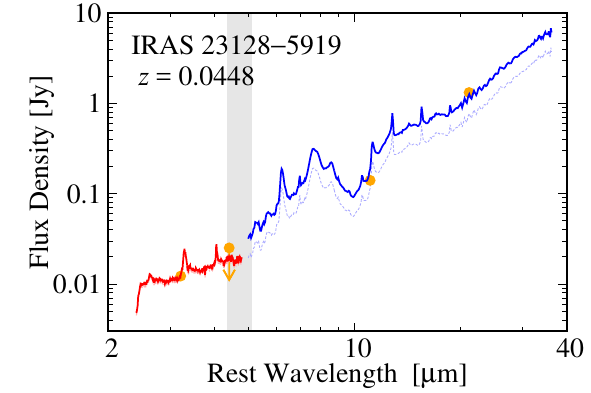}
  \end{tabular}
\caption{Combined spectrum of each \AKARI\ target. Red and blue solid curves are the \AKARI/IRC and \Spitzer/IRS spectra \ADD{scaled to match the \WISE\ photometry}, respectively \ADD{(see text)}. \ADD{Dashed lines in light colors, which are sometimes hidden behind the solid lines, are the original spectra before scaled.} Orange filled circles represent the \WISE\ photometric fluxes. The \AKARI\ and \Spitzer\ spectra are scaled so that they fit with the \WISE\ points, but the $W2$ band (4.6~\micron\ in the observed frame) flux shown with a downward arrow is used only as an upper limit. Gray shaded areas indicate the wavelength range in which the CO absorption appears.\label{fig:specAKARI}}
\end{figure*}

\ADD{After the scaling, the two spectra agreed well so that we were able to draw baselines smoothly (Section \ref{sec:cont}).
The obtained combined spectra are shown in Figure \ref{fig:specAKARI}.}
The resultant scaling shifts \ADD{from the original fluxes} were mainly within 20\%, although the shift for the \Spitzer\ spectrum of IRAS 23128$-$5919 was $+$64\%.
This large shift can be attributed to the compensation of the flux from the northern nucleus, which did not fall in the \Spitzer\ slit but blends into the \AKARI\ spectral extraction.

While the original flux uncertainties of the \AKARI\ and \Spitzer\ spectra were propagated into the combined spectra, the scale factor uncertainties were not, because only the continuum-normalized spectra were used in the analysis described below.
However, the uncertainty in the ratio between the two scale factors, which affects the determination of the longward continuum level from the \Spitzer\ spectrum, should be treated as a systematic error.
The largest uncertainty in the ratio of the two scale factors was obtained in IRAS 23128$-$5919, in which the uncertainty was $\pm$0.11 out of 1.54.
We estimate the systematic error stemming from this uncertainty in Section \ref{sec:res}.

\begin{figure*}
  \centering
  \begin{tabular}{@{}c@{}c@{}}
    \includegraphics[width=0.33\hsize]{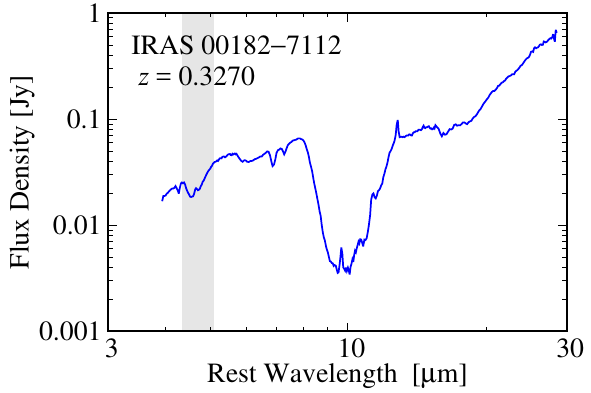} &
    \includegraphics[width=0.33\hsize]{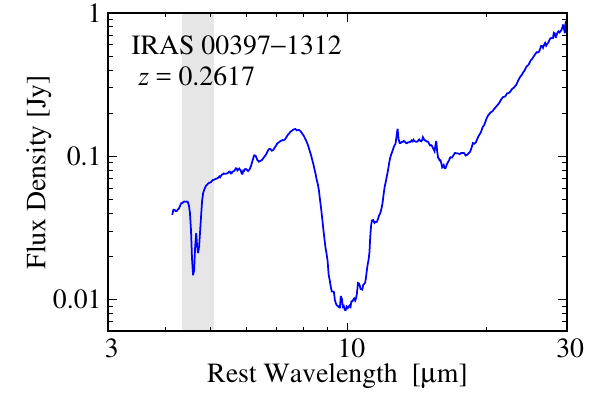} \\
    \includegraphics[width=0.33\hsize]{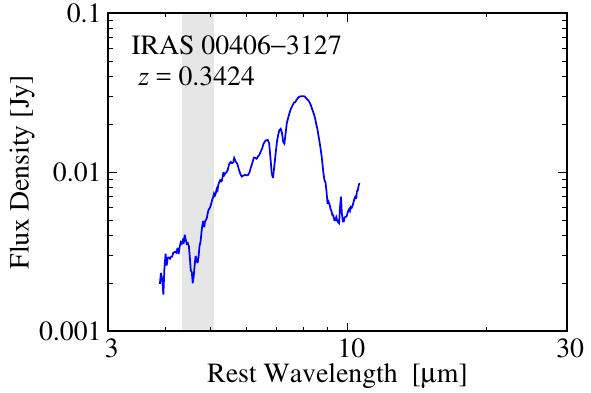} &
    \includegraphics[width=0.33\hsize]{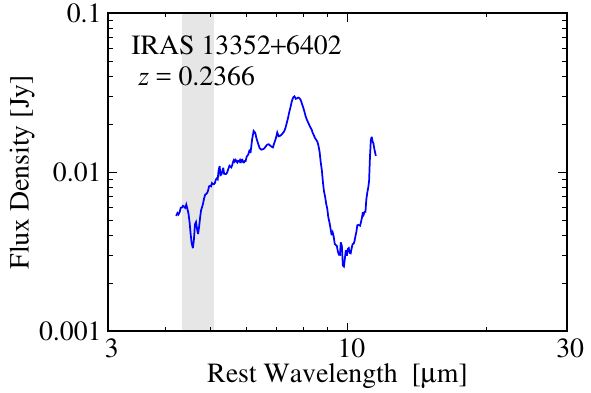} 
  \end{tabular}
\caption{\Spitzer/IRS spectrum of each \Spitzer\ target (blue solid curves). Gray shaded areas indicate the wavelength range in which CO absorption appears.\label{fig:specSpitzer}}
\end{figure*}

Figure \ref{fig:specSpitzer} shows the \Spitzer\ spectra of the four \Spitzer\ targets.
In contrast to the \AKARI\ spectra of the \AKARI\ targets, these spectra entirely cover the CO absorption within themselves.
We did not apply any scaling to these spectra because the absolute fluxes were not important, as only the continuum-normalized spectra were used in the analysis.

\section{ANALYSIS}\label{sec:ana}

\subsection{Continuum-Normalized Spectra}\label{sec:cont}

We normalized each spectrum around the CO absorption with a continuum level estimated as a cubic spline curve interpolated between the pivots at 4.15, 4.35, 5.10, and 5.40~\micron.
These pivots were taken so that they avoided the Br$\alpha$ line at 4.05~\micron, CO$_2$ absorption at 4.26~\micron, and PAH emissions at 5.27 and 5.70~\micron\ \citep{SmithEtAl_2007}.
In IRAS 13352$+$6402, we used instead a quadratic continuum that passes over the remaining three pivots, as its spectrum did not cover wavelengths shorter than 4.21~\micron.
Figure \ref{fig:continuum} shows the adopted continuum curves and the resulting normalized spectra.
In this figure, double-branched features are observed, with the branches at the long and short wavelength sides representing the $P$- and $R$-branches, respectively.
The depth of the absorption is deep, and the width of each branch is broad ($\sim0.2~\micron$) compared to that observed toward Sgr A$^*$ \citep[$\sim0.05~\micron$,][]{LutzEtAl_1996a}.
These characteristics suggest that the CO gas has a large column density and a high temperature of up to $\sim500~\mathrm{K}$.

\begin{figure*}
  \centering
  \begin{tabular}{@{}c@{}c@{}c@{}}
    \includegraphics[width=0.33\hsize]{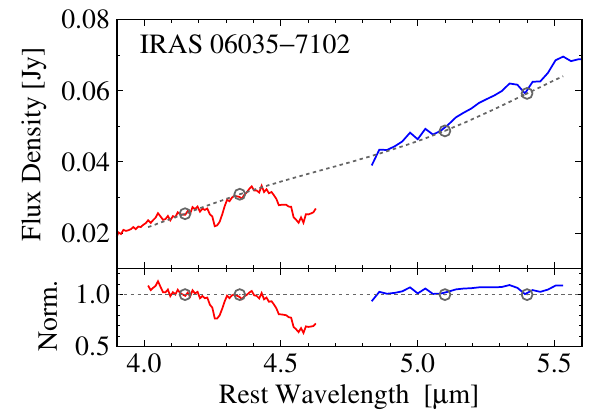}   &
    \includegraphics[width=0.33\hsize]{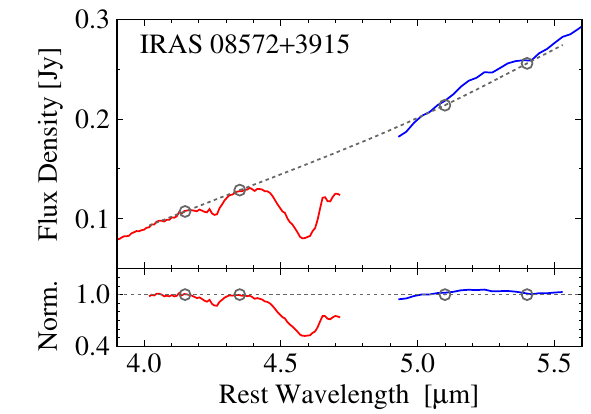}   &
    \includegraphics[width=0.33\hsize]{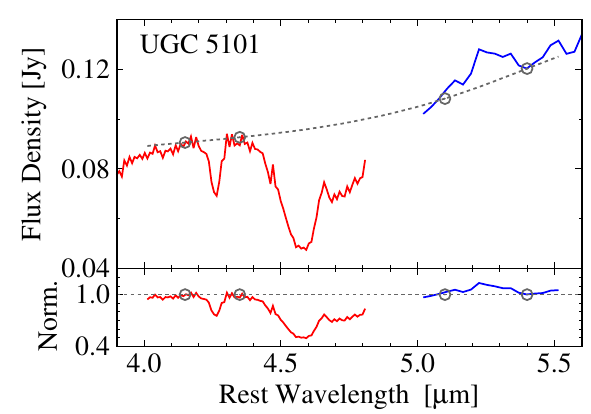}          \\
    \includegraphics[width=0.33\hsize]{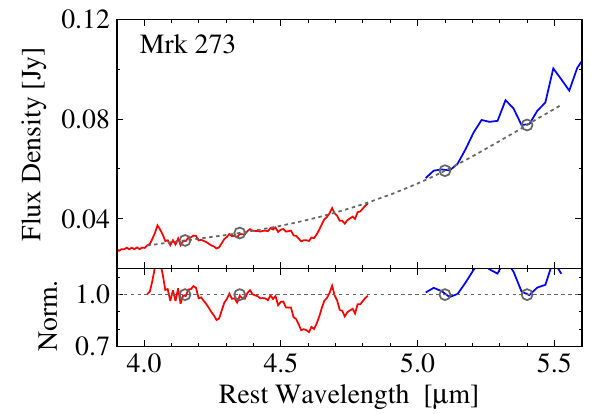}           &
    \includegraphics[width=0.33\hsize]{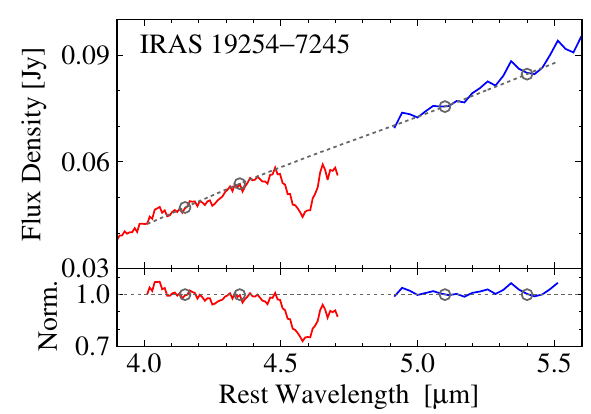}   &
    \includegraphics[width=0.33\hsize]{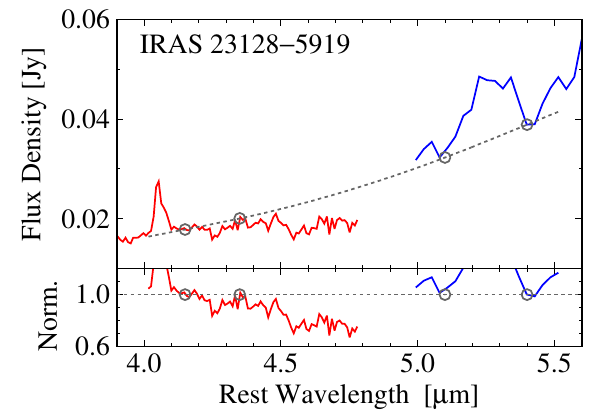} 
  \end{tabular}
\caption{Continuum curves over the CO absorption (top) and continuum-normalized spectra (bottom). Red and blue curves are the \AKARI\ and \Spitzer\ spectra, respectively. Flux uncertainty is not shown here but is indicated in Figure \ref{fig:bestfit}. The continuum spectrum of each target (gray dotted line) is taken as a cubic spline curve that passes four pivots at 4.15, 4.35, 5.10, and 5.40 \micron\ (gray open circles), except for IRAS 13352$+$6402, whose continuum is taken as a quadratic curve that passes the three pivots at longer wavelengths.\label{fig:continuum}}
\end{figure*}
\addtocounter{figure}{-1}
\begin{figure*}
  \centering
  \begin{tabular}{@{}c@{}c@{}}
    \includegraphics[width=0.33\hsize]{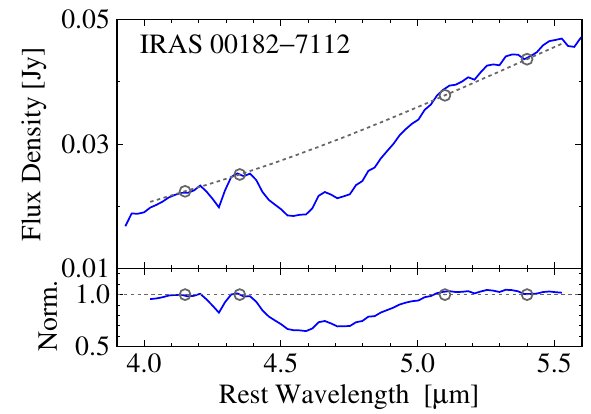}   &
    \includegraphics[width=0.33\hsize]{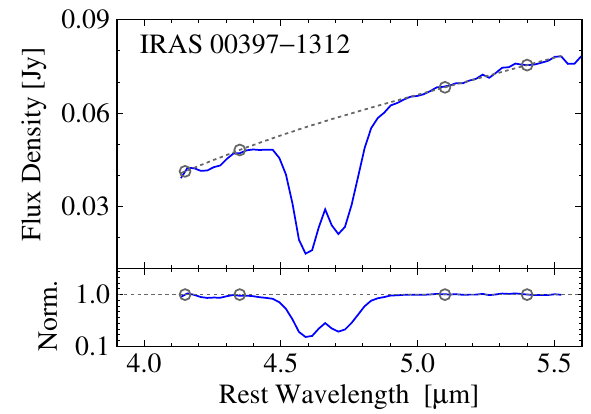}   \\
    \includegraphics[width=0.33\hsize]{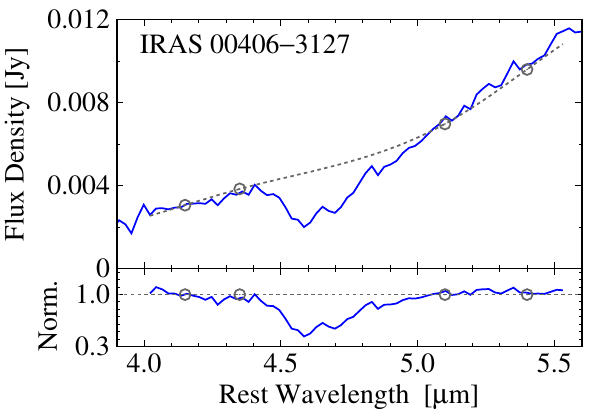}   &
    \includegraphics[width=0.33\hsize]{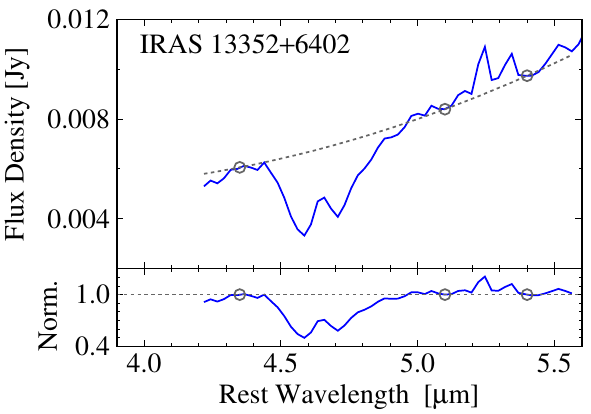}   
  \end{tabular}
\caption{\textit{(Continued)}}
\end{figure*}

\subsection{Gas-Model Fitting}

We used the plane-parallel LTE gas model developed by \citet{Cami_2002} to analyze the absorption profile.
For simplicity, we assume that the CO gas comprises of a single component with uniform number density, temperature, and turbulent velocity (velocity width).
We did not include any isotopomers other than $^{12}$C$^{16}$O.
The model gives the flux normalized to the background intensity, including the contribution of both the absorption by the gas and the thermal emission from the gas itself.
Because the observed absorption profiles suggested high gas temperatures, to accurately take the relative contribution of the gas in emission compared to the continuum into account, we needed to explicitly assume the temperature of the background radiation source. 
Assuming that the background continuum source is an optically thick hot dust sublimation layer, \ADD{we set the continuum as a blackbody} $I_{\nu,0}=B_\nu(\Tbg)$ with a sublimation temperature $\Tbg=1500~\mathrm{K}$ \citep{Barvainis_1987}.
The \ADD{intensity we observe}, $I_\nu$, is then the sum of the absorbed background, $I_{\nu,0}e^{-\ADD{\tau_\nu}}$, and the\DEL{ thermal} emission from the CO gas itself, $B_\nu(\Tco)(1-e^{-\ADD{\tau_\nu}})$, where $\ADD{\tau_\nu}$ is the optical depth of the CO gas\DEL{ at wavelength $\lambda$}, and \Tco\ is the gas temperature.
Accordingly, the continuum-normalized intensity becomes
\begin{equation}\label{eq:norm}
  I_\nu/I_{\nu,0} = e^{-\ADD{\tau_\nu}} - \frac{B_\nu(\Tco)}{B_\nu(\Tbg)}\left( 1-e^{-\ADD{\tau_\nu}} \right).
\end{equation}
\ADD{
The optical depth for a transition $(v,J)=(0,J'')\rightarrow(1,J')$ at frequency $\nu_0$ can be written as
\begin{eqnarray}
  \tau_\nu = \Nco \frac{h\nu_0}{4\pi} g_{J''} B_{J''J'} \frac{e^{-E_{J''}/k\Tco}}{Z(\Tco)} & \nonumber \\
  \times \left(1-e^{-h\nu_0/k\Tco}\right)\phi(\nu,\nu_0) &,
\end{eqnarray}
where \Nco\ is the total CO column density, $g_{J''}=(2J''+1)$ and $E_{J''}$ are the statistical weight and the energy level of the lower state, respectively, $B_{J''J'}$ is the Einstein coefficient of the transition, $Z(\Tco)$ is the partition function at \Tco, and $\phi(\nu,\nu_0)$ is a line profile.
We assumed a Gaussian profile for each transition with a common turbulent velocity \vturb:
\begin{eqnarray}
  &\phi(\nu,\nu_0)=\frac{1}{\sqrt{\pi}\Delta\nu_\mathrm{D}}e^{-(\nu-\nu_0)^2/\Delta\nu_\mathrm{D}^2},\\
  &\Delta\nu_\mathrm{D}=\frac{\vturb}{c}\nu_0.
\end{eqnarray}
From these equations, the intrinsic model spectrum can be parameterized by three variables: column density \Nco, temperature \Tco, and velocity width \vturb.
}

\ADD{We can fit the model (Equation \ref{eq:norm}) to the data using \Nco, \Tco, and \vturb\ as free parameters, after convolving it with the instrumental spectral resolutions.
The three parameters are somewhat degenerate with each other at such low resolutions since different rotational levels are not resolved, and the absorption spectrum is smoothed-out as a double-branched profile as observed in the previous subsection.}
Appendix \ref{sec:depen} explains how the parameters alter the absorption profile.

\ADD{Because most of the targets show absorption dominated spectra, the contribution from emission lines within the CO band can be generally expected to be small.
However, there is some concern.}
The Pf$\beta$ line at 4.65~\micron\ \ADD{may be} superimposed over the CO absorption.
The theoretical line ratio of Pf$\beta$/Br$\alpha$ is 0.20 under the Case B condition, with $n_\mathrm{e}\sim10^2$--$10^7~\mathrm{cm^{-3}}$ and $T_\mathrm{e}\sim$(3--30)$\times10^3~\mathrm{K}$ \citep{StoreyEtAl_1995}.
Considering the small equivalent width of the Br$\alpha$ line, we ignored the contribution from Pf$\beta$ in all the targets except for Mrk 273, in which the observed intensity near the CO band center exceeded unity.
Yano et al. (in preparation) found that this galaxy shows an anomalous Br$\beta$/Br$\alpha$ ratio in the \AKARI\ spectrum.
Accordingly, we attributed the high intensity to the effect of Pf$\beta$ and masked the data points within the $\pm3\sigma$ wavelength range around that line assuming a line width at $\mathrm{FWHM}=0.03~\micron$ and excluded those points from the fitting process.
\ADD{There is no other H\,{\footnotesize I} line predicted to be stronger than the Pf$\beta$ line.}

\ADD{
Another possible contamination is the molecular hydrogen pure-rotational line H$_2$ 0--0 S(9) at 4.69~\micron.
We estimated its flux from those of H$_2$ 0--0 $S(7)$ at 5.51~\micron\ and $S(3)$ at 9.66~\micron.
We did not use the $S(6)$, $S(5)$, and $S(4)$ lines at 6.11, 6.91, and 8.03~\micron, respectively, because they can be blended with PAH emissions. 
On the assumption that $J=5$, 9, and 11 levels are in LTE, the $S(9)$ line peak is derived to be smaller than 4\% of the continuum level at 4.69~\micron\ in all the galaxies.
We thus ignore the contribution of the $S(9)$ line.
}

\ADD{
There are some fine structure lines that possibly appear over the CO band: [Mg\,{\footnotesize IV}] 4.49~\micron, [Ar\,{\footnotesize VI}] 4.53~\micron, and [Na\,{\footnotesize VII}] 4.69~\micron.
These transitions start from high excitation stages of 80, 75, and 172~eV, respectively, and, in Seyfert galaxies, usually observed to be weaker than the [Mg\,{\footnotesize V}] 5.61~\micron\ line, whose excitation potential is 109~eV \citep{LutzEtAl_2000g,SturmEtAl_2002}.
We did not significantly detect the [Mg\,{\footnotesize V}] line above the 3$\sigma$ level in all the targets.
Hence we ignored any contribution from the three lines.
}

\section{RESULTS}\label{sec:res}

We derived the best-fit models using an iterative least-chi-square method and confirmed that their solutions definitely provided the minimum $\chi^2$ values through grid calculation.
\ADD{In the computation of the $\chi^2$ values, uncertainties in the normalized flux of the data points were taken into account.
In IRAS 00397$-$1312 and 13352$+$6402, because the observed data points appeared to be systematically displaced to shorter wavelengths from the best models we obtained at first, we manually shifted the points of the two galaxies by 0.009 and 0.010~\micron, respectively, and then again fitted the model.
These amounts are about the same as the wavelength calibration accuracy in the SL2 order of the IRS spectrum \citep[0.008~\micron\ RMS,][]{IRS_2011}. After this shifting, the reduced chi-squares for the two galaxies improved from 23.99 to 7.88 and from 7.14 to 3.39, respectively.}

\begin{figure*}
\plotone{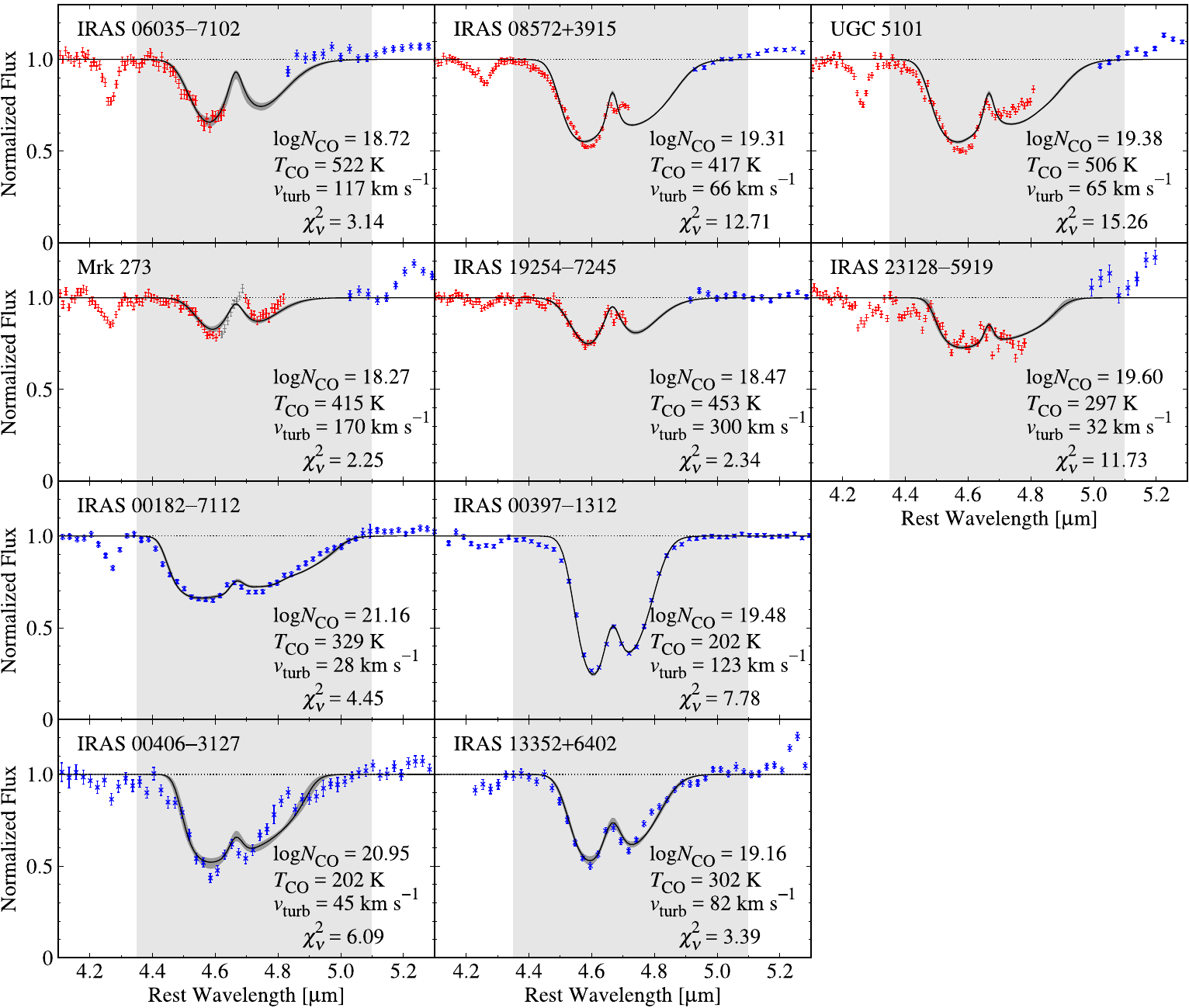} 
\caption{Results of gas-model fitting. Black solid lines denote the best-fit CO absorption profiles\ADD{, and the accompanying gray-filled curves represent the intervals corresponding to the 90\% joint confidence level}. Red and blue points are data from the \AKARI/IRC and \Spitzer/IRS spectra, respectively. Gray shaded areas indicate the wavelength range used for the spectral fitting. Dark-gray points in the spectrum of Mrk 273 were excluded from the fitting to avoid the possible contribution from the Pf$\beta$ (4.65~\micron) emission. The best-fit parameters and the goodness-of-fit $\chi^2_\nu\equiv \chi^2/\mathrm{dof}$ are noted at the right bottom corners.\label{fig:bestfit}}
\end{figure*}

Figure \ref{fig:bestfit} shows the best-fit CO gas model for each galaxy, and Table \ref{tab:bestfit} tabulates the best-fit parameters and their 90\% joint confidence ranges.
A set of color maps of the $\Delta\chi^2$ value is shown in Appendix \ref{sec:Dchi2}.
The fitting process returned quite high reduced $\chi^2$ values ($\chi^2_\nu\equiv\chi^2/\mathrm{dof}$).
We \ADD{consider} that these poor fits originate from the difficulty in determining the continuum levels over the broad CO band.
\ADD{Some other possibilities concerning limitations of the current model are discussed in Section \ref{sec:beyond}.}
We speculate that there is a systematic trend in which the observed flux in the red wing of the $P$-branch is higher than the model prediction; possible origins of this trend are \ADD{also discussed in Section \ref{sec:beyond}}.
The CO absorptions in IRAS 00182$-$7112 and IRAS 08572$+$3915 were also analyzed by \citetalias{SpoonEtAl_2004b} and \citetalias{ShirahataEtAl_2013}, respectively.
Comparisons between their results and ours appear in Appendix \ref{sec:previous}.

\begin{deluxetable*}{cllcr} 
\tablecaption{Best-Fit parameters and goodness of fit\label{tab:bestfit}}
\tablewidth{0pt}
\tablehead{
\colhead{Object} & \colhead{$\log \Nco$} & \colhead{\Tco} & \colhead{\vturb}  & \colhead{$\chi^2/{\rm dof}$}\\
\colhead{}       & \colhead{(in \psqcm)} & \colhead{(K)}  & \colhead{(\kmps)} & \colhead{}
}
\startdata
IRAS 06035$-$7102 & $18.72^{+0.25}_{-0.08}$ & $522^{+ 71}_{-104}$ & $>49            $ & $119.5 / 38$ \\
IRAS 08572$+$3915 & $19.31^{+0.09}_{-0.08}$ & $417^{+ 23}_{- 23}$ & $ 66^{+ 5}_{- 4}$ & $546.3 / 43$ \\
UGC 5101          & $19.38^{+0.10}_{-0.09}$ & $506^{+ 35}_{- 34}$ & $ 65^{+ 5}_{- 5}$ & $747.9 / 49$ \\
Mrk 273           & $18.27^{+0.07}_{-0.04}$ & $415^{+ 76}_{- 63}$ & $>52            $ & $ 94.5 / 42$ \\
IRAS 19254$-$7245 & $18.47^{+0.09}_{-0.02}$ & $453^{+ 37}_{- 42}$ & $>62            $ & $103.0 / 44$ \\
IRAS 23128$-$5919 & $19.60^{+2.61}_{-0.32}$ & $297^{+ 66}_{-169}$ & $ 32^{+ 5}_{-11}$ & $563.3 / 48$ \\
IRAS 00182$-$7112 & $21.16^{+0.70}_{-1.06}$ & $329^{+124}_{- 55}$ & $ 28^{+ 7}_{- 2}$ & $133.6 / 30$ \\
IRAS 00397$-$1312 & $\ADD{19.48}^{+\ADD{0.08}}_{-\ADD{0.07}}$ & $\ADD{202}^{+\ADD{  9}}_{-\ADD{  9}}$ & $\ADD{123}^{+\ADD{ 9}}_{-\ADD{ 8}}$ & $\ADD{217.8 / 29}$ \\
IRAS 00406$-$3127 & $20.95^{+1.30}_{-1.37}$ & $202^{+130}_{- 57}$ & $ 45^{+24}_{- 7}$ & $182.7 / 30$ \\
IRAS 13352$+$6402 & $\ADD{19.16}^{+\ADD{0.23}}_{-\ADD{0.15}}$ & $\ADD{302}^{+\ADD{ 36}}_{-\ADD{ 40}}$ & $\ADD{ 82}^{+\ADD{24}}_{-\ADD{16}}$ & $\ADD{ 94.8 / 28}$ 
\enddata
\tablecomments{All errors are quoted at the 90\% joint confidence level for three parameters ($\Delta\chi^2=6.25$), instead of for one parameter of interest. Lower limits for \vturb\ are presented if it cannot be constrained within the parameter range ($<300~\kmps$) at the 99\% joint confidence level.}
\end{deluxetable*}

We found that the observed CO gas has a high temperature and large column density.
The average \Tco\ is 360~K, which is far higher than the typical temperature of molecular gas in ordinary star-forming regions \citep[10--$10^2~\mathrm{K}$,][]{HollenbachEtAl_1999}.
The logarithm of \Nco\ in units of \psqcm\ is on average 19.5,\footnote{Throughout this paper, the logarithm of column density is presented in units of \psqcm.} which corresponds to a molecular hydrogen column density of $\ADD{\log N_\mathrm{H_2}\sim23.5}$ if \ADD{a CO abundance $[\mathrm{CO}]/[\mathrm{H_2}]=10^{-4}$ is adopted}.
This large column density can be converted to extinction at $M$-band (5~\micron) of $\ADD{A_M\sim8}$ if we assume $\ADD{\Nh=2N_\mathrm{H_2}}$ \ADD{and} Galactic relations $\Nh/A_V=1.9\times10^{21}~\psqcm~\mathrm{mag^{-1}}$ \citep{BohlinEtAl_1978} and $A_V/A_M\sim40$ \citep[$R_V=3.1$,][]{Draine_2003}.
Because we were able to observe CO gas absorption with this large column density, the extinction by dust must not be too high.
This could suggest that the dust-to-gas ratio in AGN neighborhoods is lower than the Galactic value.
\ADD{Note that the estimate of $A_M$ have a major uncertainty depending on the assumption of the abundance ratio.
The ratio measured in Galactic objects differs in (0.8--3)$\times 10^{-4}$ \citep{Dickman_1978,FrerkingEtAl_1982,WatsonEtAl_1985,BlackEtAl_1990,LacyEtAl_1994}.
There is no direct measurement of the CO abundance in ULIRGs, to our knowledge.
The range of the abundance leads $3<A_M<10$.}

For the results of the \AKARI\ targets, we estimated the systematic error stemming from the scaling of the \AKARI\ and \Spitzer\ spectra.
In IRAS 23128$-$5919, the ratio of the scaling factor for the \Spitzer\ spectrum to that for \AKARI\ was determined to be $1.54\pm0.11$; this uncertainty was the largest among all the \AKARI\ targets (Section \ref{sec:obs}).
Changing the amount of scaling by this uncertainty, the fitting result of this galaxy differs by $\Delta(\log\Nco)\sim0.2$, $\Delta\Tco\sim40$~K, $\Delta\vturb\sim2~\kmps$.
We therefore estimate that the systematic error is smaller than these values and conclude that this error does not affect the following discussion.

\section{DISCUSSION}\label{sec:dis}

\ADD{
\subsection{Limitations of the Current Model}\label{sec:beyond}
}

As the model simplifies the CO gas temperature to a uniform value, the results should be treated as first order approximations.
Here, we note some effects that are not incorporated in the current approximation.

If there are two or more temperature components, the higher-temperature components will make the column density appear to be larger.
Under the current spectral fitting, the band wings strictly constrain the solution.
The higher-temperature components work to lower the flux in the band wings in an effect similar to that of increasing the column density (see also Appendix \ref{sec:depen}).
The high reduced $\chi^2$ values we obtained in the fitting may therefore be the result of over simplification of multi-temperature components.
Although this factor is of concern, given the quality of the \AKARI\ and \Spitzer\ spectra, it is difficult to fit a more complicated multi-temperature gas model.

\begin{figure}
\plotone{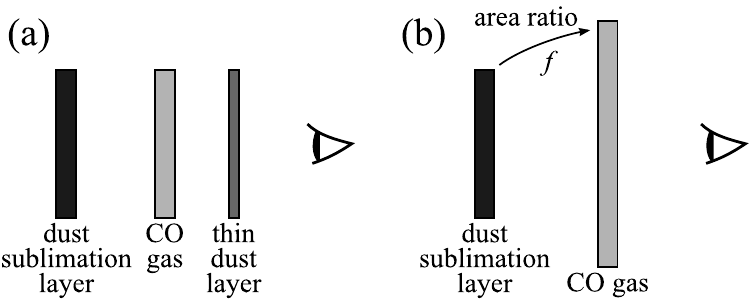} 
\caption{Two types of geometries beyond the model assumption. (a) Geometry in which another foreground thin dust layer is present. (b) Geometry in which the CO gas layer extends beyond the background dust sublimation layer with an area ratio $f$.\label{fig:beyond}}
\end{figure}

If there is an optically-thin foreground dust screen in front of the CO gas, as shown Figure \ref{fig:beyond}(a), its thermal radiation will fill up the absorption.
The observed potential trend in the shallow red wing of the $P$-branch (Section \ref{sec:res}) may be caused by this effect.
Similarly, if the CO gas extends to a larger solid angle than the continuum source as shown in Figure \ref{fig:beyond}(b), the band profile would be skewed, as the outer part of the CO gas contributes merely by emission without absorbing the background light.
This effect can be quantified in terms of the area ratio, $f$, of the CO gas to the background source.
We can introduce $f$ into Equation (\ref{eq:norm}) by multiplying it with the second term.
The increase of the normalized flux density owing to $f$ is larger at longer wavelengths because $B_\nu(\Tco)$ is redder than $B_\nu(\Tbg)$ at 4.67~\micron.
This effect can also cause the systematic trend seen in the shallow red wing in the $P$-branch.
However, if $\Tco\sim300~\mathrm{K}$, $f$ must be very large ($\sim10^3$) because $B_\nu(\Tco)$ is at the Wien side at that wavelength.
This effect would therefore appear not to fully explain the systematic trend.

We did not adopt any Doppler shifts into the modeled absorption.
The implied systematic poor fit in the red wing in the $P$-branch can also be interpreted as blueshift from the systemic velocity.
A shift of $\sim0.02~\micron$ requires a velocity of $\sim-1000~\kmps$.
Such high-velocity molecular outflows have in fact been found in far-IR OH emission lines \citep[e.g.,][]{FischerEtAl_2010d,SpoonEtAl_2013} and are comparable with column densities as large as $\Nh\sim10^{23}~\psqcm$ \citep{Gonzalez-AlfonsoEtAl_2017}.
Thus, the poor fit in the $P$-branch may be a signature of warm molecular outflows.
However, such a conclusion from only the present broad blurred absorption profiles is not robust, and verification will require other observations at higher spectral resolutions that can resolve different rotational levels.

Some other absorption bands possibly overlap \ADD{the CO band}.
If present, $^{13}$CO gas and CO ice features appear at wavelengths longer than the $^{12}$CO gas band center.
These would deepen the $P$-branch of the $^{12}$CO profile, but this contribution is contrary to the observed pattern, suggesting that there is no signature of such features.
On the other hand, the XCN ice feature, whose center is 4.62~\micron\ and FWHM is 0.05~\micron, possibly superimposes on the $R-$branch of the gas phase $^{12}$CO band, as was observed in the starburst/AGN galaxy NGC 4945 \citep{SpoonEtAl_2003} and the starburst galaxy NGC 253 \citep{YamagishiEtAl_2011}.
However, this ice band is narrower than the observed width of the CO $R$-branch and thus unlikely to mimic it.
Although there are weak implications of such an additional narrow absorption at the peak of the branch in some targets such as UGC 5101 (Figure \ref{fig:bestfit}), we cannot rule out the possibility that the responsible absorber is not XCN ice but another, colder CO gas component.

As it is difficult to discuss the above effects in detail based on the present \AKARI\ and \Spitzer\ spectra, we do not further pursue them in this paper.

\subsection{Heating Mechanism}

Our most important finding is that the observed CO gas has a high temperature and a large column density.
The typical temperature $\Tco\sim400~\mathrm{K}$ is far higher than that of molecular gas in ordinary star-forming regions \citep[10--$10^2~\mathrm{K}$,][]{HollenbachEtAl_1999}.
The typical column density $\Nco\sim10^{19}~\psqcm$ corresponds to $N_\mathrm{H_2}\sim10^{23}~\psqcm$ if we assume an abundance ratio $[\mathrm{CO}]/[\mathrm{H_2}]=2\times10^{-4}$ \citep{Dickman_1978}.
Here, we consider what mechanism can heat the observed large columns of warm gas through a discussion of three candidates.

The first candidate heating source is ultraviolet (UV) photons emitted from the central accretion disk.
An intense UV radiation field incident on a cloud forms a photon-dominated region (PDR) \ADD{and} determines its thermal and chemical structures.
Several authors have modeled \ADD{the PDR under} various conditions \citep[e.g.,][]{TielensEtAl_1985,HollenbachEtAl_1999}.
\ADD{\citet{MeijerinkEtAl_2005a} developed PDR models at four conditions specified by the combination of the gas number density ($n_\mathrm{H}=10^3$ or $10^{5.5}~\mathrm{cm^{-3}}$) and the incident far-UV flux ($G_0=10^3$ or $10^5$).
Here $G_0$ is a flux measure normalized at $1.6\times10^{-3}~\mathrm{erg~cm^{-2}~s^{-1}}$ \citep{Habing_1969}.
In all the cases except for the high-density and low-flux case ($n_\mathrm{H}=10^{5.5}~\mathrm{cm^{-3}}$ and $G_0=10^3$), the maximum gas temperature exceeds $10^3~\mathrm{K}$, which is sufficiently higher than the observed values of \Tco.
However, in the models, the gas temperature afterward decreases steeply because of strong attenuation by dust and drops to $10^2~\mathrm{K}$ before the column density \Nh\ reaches $10^{22}~\psqcm$.
Moreover, the CO abundance is suppressed in this PDR region due to photo-dissociation: the CO/C ratio is lower than unity at $\Nh<10^{22}~\psqcm$.
The achievable warm ($>10^2~\mathrm{K}$) CO column density is only $\Nco\sim10^{16}~\psqcm$ and far smaller than the observed values ($\gtrsim10^{18}~\psqcm$) by two orders of magnitude.
This suggests that UV heating cannot represent the observed large columns of warm gas.}

In the study of CO pure rotational emission lines, merely detecting high-$J$ lines with a high excitation temperature does not rule out the possibility of PDRs as the origin of the emission \citep{MashianEtAl_2015}.
Actually, \citet{LoenenEtAl_2010} reproduced the spectral line energy distribution (SLED) of the starburst galaxy M82 up to $J=12$ using only PDRs.
However, the PDRs introduced in their analysis had column densities only on the order of $\Nh\sim10^{21}~\psqcm$.
In our observation of the fundamental CO ro-vibrational transition, the broad band width was equivalent to the detection of highly excited lines at $J\gtrsim20$.
More importantly, the obtained column density is too large ($\Nh\gtrsim10^{22}~\psqcm$) to be reproduced by PDRs and requires other heating sources.

The next candidate is shock heating.
Given that many of the observed galaxies are merging systems or in disturbed morphologies, it is reasonable to conjecture that their gas is powered by shocks arising from turbulent motion.
A number of studies have discussed the physical and chemical processes in shock propagation in interstellar clouds \citep[e.g.,][]{McKeeEtAl_1984,HollenbachEtAl_1989,NeufeldEtAl_1989}.
With a shock velocity of $\sim100~\kmps$ and a pre-shock density of $10^4$--$10^6~\mathrm{cm^{-3}}$, the post-shock temperature structure begins with the initial value of $\ga10^4~\mathrm{K}$, where the gas is heated by UV photons from the vicinity of the shock front.
The temperature then gradually decreases until the UV radiation is sufficiently attenuated and molecular recombination becomes effective.
After this point, OH molecules become the major coolant and the gas temperature rapidly falls below $\sim10^3~\mathrm{K}$.
This rapid drop is followed by an equilibrium between OH cooling and heating owing to $\mathrm{H_2}$ formation on dust grains.
CO rotational transitions also significantly contribute to cooling in this phase.
Under this balance, the temperature remains at $\sim10^2~\mathrm{K}$ until $\mathrm{H_2}$ formation is completed.

Because the molecular-forming region downstream of a shock provides gas temperatures higher than $10^2~\mathrm{K}$, shock heating can explain observed \Tco\ values.
On the other hand, the scale of the column density in the warm gas layer is $\Nh\la10^{22}~\psqcm$ and the CO relative abundance is nearly constant at $\sim10^{-4}$ \citep{HollenbachEtAl_1989,NeufeldEtAl_1989}.
Hence, shock heating can produce a warm CO gas with column densities of up to $\Nco\sim10^{18}~\psqcm$.
This upper bound is comparable to the smallest \Nco\ we observed.
Thus, while there is still room for a partial contribution from shock heating, it cannot fully account for the heating mechanism of the CO gas.

The third heating source candidate is X-ray photons emitted from the nuclear region of the AGN.
A strong X-ray radiation field incident on a cloud creates an X-ray-dominated region (XDR) that drives its internal thermal and chemical processes.
Although this process is similar to PDR formation by UV photons, X-ray photons can penetrate more deeply into clouds because of their small cross sections.
\citet{MeijerinkEtAl_2005a} modeled XDRs with chemical reactions under four conditions involving combinations of two densities ($n_\mathrm{H}=10^3$ and $10^{5.5}~\mathrm{cm^{-3}}$) and two incident X-ray fluxes ($F_\mathrm{X}=1.6$ and $160~\mathrm{erg~cm^{-2}~s^{-1}}$).
In all cases except those with high density and low flux, the simulated temperatures are in the range $10^2$--$10^4~\mathrm{K}$ and accord with observed \Tco\ values.
Moreover, the temperatures remain higher than $10^2~\mathrm{K}$ up to a column density of $\Nh\sim10^{24}~\psqcm$.
This scale of heating is more than two orders of magnitude larger than those resulting from the previous two mechanisms.
In all cases, \Nco\ reaches $10^{18}~\psqcm$ before the temperature fall to $10^2~\mathrm{K}$ \citep[compare Figures 6 and 8 in][]{MeijerinkEtAl_2005a}, demonstrating that such XDRs can account for the observed warm CO gas of a large column density.
In addition, the dense and intense case ($n_\mathrm{H}=10^{5.5}~\mathrm{cm^{-3}}$ and $F_\mathrm{X}=160~\mathrm{erg~cm^{-2}~s^{-1}}$) gives the largest \Nco\ at $T=100~\mathrm{K}$ among all cases ($10^{20}~\psqcm$), suggesting that the two sources of large $\Nco\sim10^{21}~\psqcm$, IRAS 00182$-$7112 and IRAS 00406$-$3127, require more extreme XRDs.

We attempted to distinguish shock and X-ray heating on the basis of the CO SLED of the rotational transitions.
Such lines are one of the major coolants in a shocked gas.
This discussion is analogous to that argued by \citet{MeijerinkEtAl_2013} for NGC 6240 and Mrk 231, which are nearby merger/starburst and Sy1 galaxies, respectively.
These two galaxies are similar in terms of their CO SLED shapes,with a flat distribution up to higher rotational levels, but differ in terms of their line-to-continuum ratios $\Lco/\LIR$, where \Lco\ is the total luminosity of the $^{12}\mathrm{CO}$ lines including CO $J=1$--0 through $J=13$--12.
While the ratio in Mrk 231 is $\sim6\times10^{-5}$ \citep{vanEtAl_2010e}, that in NGC 6240 is $\sim7\times10^{-4}$, which is approximately an order of magnitude higher than the former value \citep{MeijerinkEtAl_2013}.
The authors concluded that this large difference results from the difference in gas heating processes in the two galaxies.
UV and X-ray photons effectively heat both gas and dust, and resulting in a far-IR spectrum that is continuum-dominated and, consequently, a small line-to-continuum ratio.
The authors also predicted the maximum line-to-continuum ratio in PDRs and XDRs as $\sim10^{-4}$.
Shocks, on the other hand, selectively heat gas by compression, maintaining the thermal decoupling of dust and attaining ratios higher than $10^{-4}$.
This leads to the conclusion that the dominant power sources for the CO gas clouds emitting rotational lines in Mrk 231 and NGC 6240 are UV to X-ray photons and shocks, respectively.

To make a similar analysis, we referred to \citet{PearsonEtAl_2016}, who observed the CO emission lines from $J=5$--4 to $J=13$--12 in 43 ULIRGs using the \textit{Herschel Space Observatory}.
Seven of their sample galaxies are included in our targets, namely, all of the \AKARI\ targets and IRAS 00397$-$1312.
We calculated the sub-total CO luminosity $\Lco^\prime$ spanning from $J=5$--4 to $J=13$--12 in the seven objects and then evaluated the line-to-continuum ratio from $\Lco^\prime/\LIR$.
We also derived the ratios in NGC 6240\footnote{Note that the line luminosities in NGC 6240 measured by \citet{PearsonEtAl_2016} were substantially lower than the measurements by \citet{MeijerinkEtAl_2013}. \citet{PearsonEtAl_2016} claimed that the reason for this is unclear, as the two papers had used the same observation.} and Mrk 231 for comparison and found the values of $4\times10^{-4}$ and $6\times10^{-5}$, respectively.
Among the seven galaxies in our sample, the lowest ratio was found in IRAS 08572$+$3915, in which the ratio is $3\times10^{-5}$, which is even smaller than that in Mrk 231.
This object also shows a continuum dominated spectrum in the near- and mid-IR region.
These facts strongly rule out the possibility of powerful shock heating within the galaxy.
On the other hand, the highest ratio was found in IRAS 06035$-$7102, where the ratio is $2\times10^{-4}$, suggesting the possibility of shock heating, a conclusion supported by the facts that: a) the \Nco\ value found in this galaxy is relatively smaller than that in the other targets, and b): a widely used shock tracer $\mathrm{H_2}$ vibrational emission around 2~\micron\ was detected in this source \citep{DannerbauerEtAl_2005}.
However, as the ratio $\Lco^\prime/\LIR$ is lower by a factor of two than that in NGC 6240, shocks in IRAS 06035$-$7102 would not be as energetic as those in NGC 6240.
In the remaining four galaxies, the ratio ranged within (0.8--1.0)$\times10^{-4}$, and there was no evidence of shock heating in these objects.

Given the above discussion, we conclude that the most reasonable heating mechanism of observed warm CO gas with large column densities is X-ray heating, which leads to the further conclusion that the observed CO gas is in the vicinity of the nucleus.
Although the possibility that shock heating accounts for a substantial fraction of the total power cannot be ruled out in some objects, its contribution must be smaller than that of X-ray photons.
In the following discussion, we presuppose that the primal heating source of the CO gas observed in absorption is X-ray photons from the central region of the AGN.

\subsection{The Relations between the Column Density, Temperature, and IR Luminosity}

\begin{figure}
\plotone{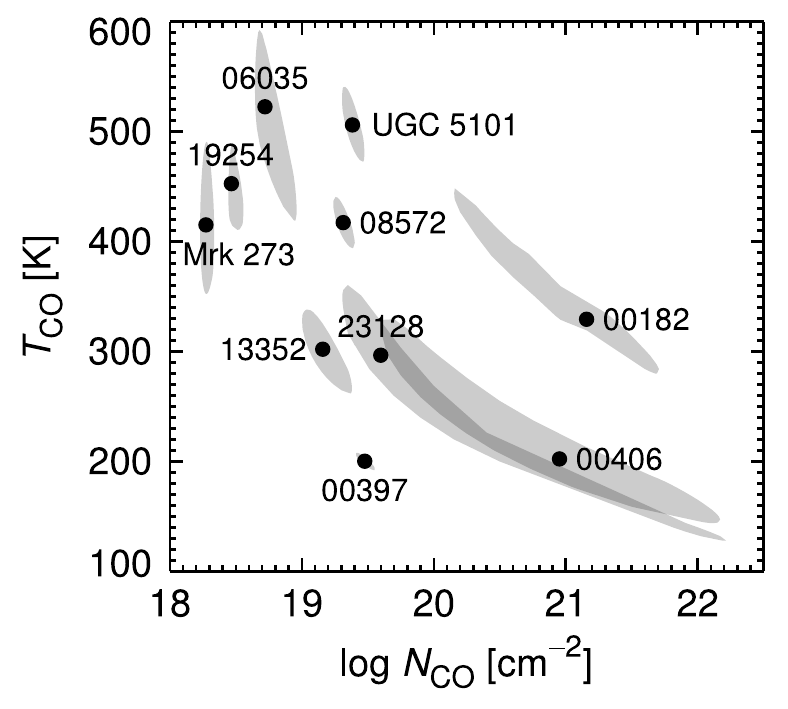}
\caption{Plot of \Tco\ versus \Nco. Black points indicate the best-fit values. Gray shaded areas are the projection of the three-dimensional 90\% joint confidence regions along the \vturb\ axis. The unprojected 3D confidence regions are shown in Figure \ref{fig:chi2map}.\label{fig:NcoVsTco}}
\end{figure}

\begin{figure}
\plotone{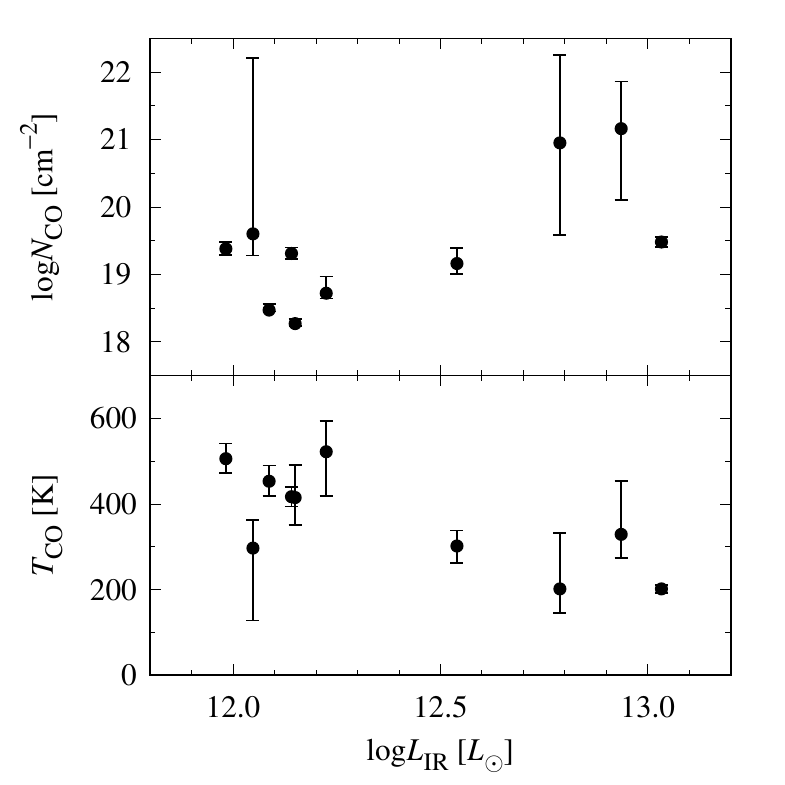} 
\caption{Dependence of \Nco\ (top) and \Tco\ (bottom) on \LIR. The abscissa is the logarithm of \LIR\ in units of the solar luminosity $L_\sun$.\label{fig:VsLIR}}
\end{figure}

Figure \ref{fig:NcoVsTco} plots \Tco\ versus \Nco, showing that \Tco\ decreases with \Nco.
Although the confidence ranges of the best-fits show a strong degeneracy between the two parameters, these ranges are smaller than the scale of the overall decreasing tendency.\DEL{
We assume here that this trend is not an arithmetic artifact but an intrinsic nature of the CO gas around AGNs.}
\ADD{This result} can be interpreted as an attenuation in which the gas a large-column distant from the heating source is heated less than gas near the source in a small column if the heating powers in the respective targets are nearly the same.
Figure \ref{fig:VsLIR} shows that the dependence of \Nco\ and \Tco\ on the IR luminosity, \LIR, is not clear.
This supports the interpretation that the heating process in the CO gas observed in absorption is a local phenomenon occurring near the power source and is not tightly related to the global activities of the host galaxy.

\subsection{Comparison with X-Ray Observations}

To clarify the location of the region in which CO absorption originates, we compared the obtained CO column densities with neutral column densities estimated from other X-ray observations.
\citet{BrightmanEtAl_2011a} performed a systematic X-ray spectral analysis for the \textit{XMM-Newton} data (0.2--10~keV) of 126 nearby galaxies and investigated their AGN properties using a spectral model that assumes a spherical toroidal obscuring material.
Four objects in their sample are common with our targets.
Table \ref{tab:Xobs} summarizes their line-of-sight hydrogen column densities determined by the X-ray spectral analysis.

\begin{deluxetable}{cc} 
\tablecaption{Column density inferred from X-ray observations\label{tab:Xobs}}
\tablewidth{0pt}
\tablehead{
\colhead{Object} & \colhead{\Nhx} \\ 
\colhead{} & \colhead{($10^{22}~\psqcm$)} 
}
\startdata
UGC 5101          & $49.6^{+25.4}_{-18.2}$ \\ 
Mrk 273           & $59.7^{+17.1}_{-12.8}$ \\ 
IRAS 19254$-$7245 & $38.1^{+39.2}_{-21.7}$ \\ 
IRAS 23128$-$5919 & $>150                $ 
\enddata
\tablecomments{
Column 1: object name.
Column 2: hydrogen absorption column density, with errors quoted at 90\% confidence level for one parameter of interest \citep{BrightmanEtAl_2011a}.
}
\end{deluxetable}

\begin{figure}
\plotone{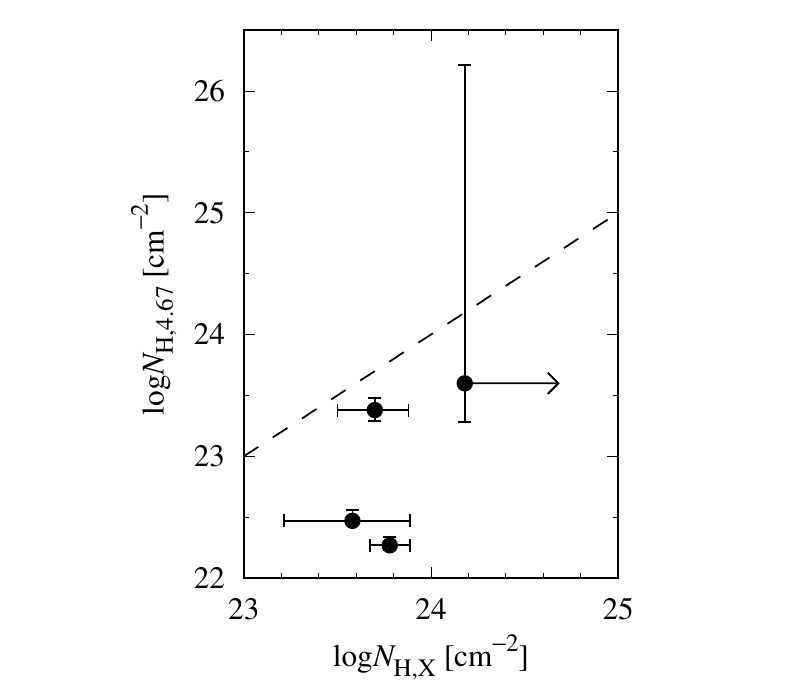} 
\caption{Comparison of the hydrogen column densities derived from an X-ray spectral analysis (abscissa; Table \ref{tab:Xobs}) and from CO absorption (ordinate). The dashed line denotes the identity.\label{fig:NHXVsNco}}
\end{figure}

We converted \Nco\ into \Nh\ assuming a ratio of $\Nco/\Nh\sim10^{-4}$ for the XDRs in the high-density cases of \citet{MeijerinkEtAl_2005a}.
Note that \Nco/\Nh\ is not constant if the incident X-ray field is intense, as the CO abundance is suppressed at small \Nh; here, we ignore this dependence for simplicity.
Figure \ref{fig:NHXVsNco} compares the two types of hydrogen column density.
The hydrogen column density derived from the CO absorption, \Nhco, is 2--30 times smaller than that inferred from the X-ray spectral analysis, \Nhx.
This comparison indicates that the two columns trace the amount of gas at different depths.
The X-ray-derived column density \Nhx\ measures the gas in front of the central AGN nucleus using the X-ray radiation from it as the background.
By contrast, \Nhco\ should measure the gas outside the X-ray emitting region, tracing a smaller amount of foreground gas.
This consideration is consistent with the assumption that the near-IR background continuum source for CO absorption is a region in front of the nucleus being warmed by the radiation from the nucleus.
Therefore, we conclude that the CO absorption originates in molecular gas distributed outside the X-ray emitting region.

\subsection{Comparison with the 9.7~\micron\ Silicate Absorption}

\begin{figure}
\plotone{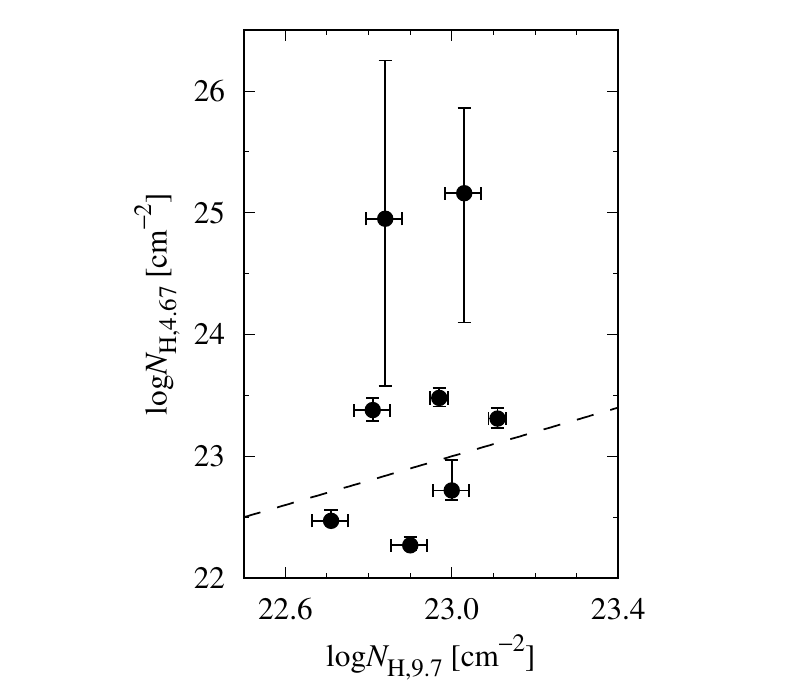} 
\caption{Comparison of hydrogen column densities derived from the 9.7~\micron\ silicate dust absorption strength (abscissa; Table \ref{tab:tausi}) and from the CO absorption (ordinate). The dashed line denotes the identity.\label{fig:NHsiVsNco}}
\end{figure}

Another major indicator of the degree of obscuration is the strength of the 9.7~\micron\ silicate dust feature, as is seen in Figures \ref{fig:specAKARI} and \ref{fig:specSpitzer}.
The optical depth of the feature $\tau_{9.7}$ in nearby AGNs have been measured in several studies \citep[e.g.,][]{DartoisEtAl_2007,ImanishiEtAl_2007,Imanishi_2009}.
The depth $\tau_{9.7}$ can be converted into the hydrogen column density using two relations connected to visual extinction $A_V$: $A_V/\tau_{9.7}=18.0~\mathrm{mag}$ \citep{Whittet_2003}, and $\Nh/A_V=1.9\times10^{21}~\psqcm~\mathrm{mag^{-1}}$ \citep{BohlinEtAl_1978}.
Table \ref{tab:tausi} cites the values of $\tau_{9.7}$ from other papers and tabulates the derived column density, \Nhsi.
Figure \ref{fig:NHsiVsNco} compares the two types of hydrogen column density, \Nhco\ and \Nhsi.
In contrast to the comparison with the X-ray observations, \Nhco\ is similar to or a bit larger than \Nhsi.
We conclude that CO-absorbing gas and silicate dust roughly coexist in the same region.

\begin{deluxetable}{ccc}
\tablecaption{9.7~\micron\ silicate feature from literature\label{tab:tausi}}
\tablewidth{0pt}
\tablehead{
\colhead{Object} & \colhead{$\tau_{9.7}$} & \colhead{\Nhsi}\\
\colhead{} & \colhead{} & \colhead{($10^{22}~\psqcm$)}
}
\startdata
IRAS 06035$-$7102 & 2.9 (10\%)\tablenotemark{a} &  9.9 \\
IRAS 08572$+$3915 & 3.8  (5\%)\tablenotemark{b} & 13   \\
UGC 5101          & 1.9 (10\%)\tablenotemark{a} &  6.5 \\
Mrk 273           & 2.3 (10\%)\tablenotemark{a} &  7.9 \\
IRAS 19254$-$7245 & 1.5 (10\%)\tablenotemark{a} &  5.1 \\
IRAS 00182$-$7112 & 3.1 (10\%)\tablenotemark{a} & 11   \\
IRAS 00397$-$1312 & 2.7  (5\%)\tablenotemark{c} &  9.2 \\
IRAS 00406$-$3127 & 2.0 (10\%)\tablenotemark{a} &  6.8 
\enddata
\tablecomments{Column 1: object name. Column 2: optical depth of the 9.7~\micron\ silicate dust absorption with uncertainty in parenthesis. Column 3: hydrogen column density calculated from $\tau_{9.7}$ (see the text).}
\tablenotetext{a}{\citet{DartoisEtAl_2007}.}
\tablenotetext{b}{\citet{ImanishiEtAl_2007}.}
\tablenotetext{c}{\citet{Imanishi_2009}.} 
\end{deluxetable}

\subsection{Distribution of the CO Gas}

In our model, we assume that the covering factor of CO gas is unity.
This assumption is not trivial but strongly supported by the observed deep absorption profiles.
Such absorption indicates that the continuum source is almost entirely covered by the foreground gas.
Because molecular clouds in star-forming regions distribute randomly, they cannot explain a large covering factor close to unity.
Thus, we assume that the absorber is a molecular cloud just in front of the continuum source.

\begin{figure}
\plotone{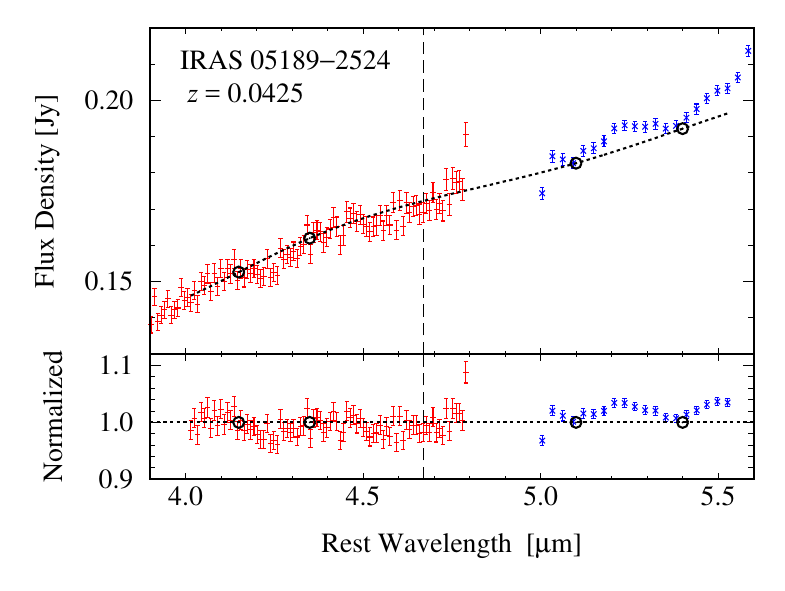} 
\caption{Top: 4.0--5.5~\micron\ spectrum of the Sy2 galaxy IRAS 05189$-$2524. Red and blue points represent the \AKARI\ and \Spitzer\ data, observation ID 1100129.1 and AOR key 16909568, respectively. Data reduction was performed in the manner described in Section \ref{sec:obs}. The black dotted line denotes the continuum curve determined as explained in Section \ref{sec:cont}. The black dashed line indicates the expected wavelength of the CO band center. Bottom: continuum-normalized spectrum of the same galaxy.\label{fig:IR05}}
\end{figure}

It is remarkable that a significant fraction of Sy2 galaxies do not show CO absorption.
In the \AKARI\ program AGNUL, \ADD{IRAS 05189$-$2524, which is classified as Sy2 or hidden broad-line Sy1 \citep{Veron-CettyEtAl_2006},} does not show any signature of CO absorption, as shown in Figure \ref{fig:IR05}.
\citet{LutzEtAl_2004} searched \textit{ISO} spectra of nearby 19 type-1 and 12 type-2 AGNs for CO absorption, but none showed the signature of the absorption.
One of their targets, the famous Sy2 galaxy NGC 1068, was re-observed by \citet{GeballeEtAl_2009} using UKIRT, with CO absorption once again undetected.
Given the implications of the comparison with X-ray observations, the CO absorption is likely to trace the region outside the X-ray emitting region.
The close-to-unity covering factor suggests that the CO absorption originates near the dust sublimation layer at the inner rim of the obscuring material surrounding the nucleus.
Here we speculate that the presence or absence of CO absorption originates from a complex innermost geometry in the putative AGN torus, e.g., a concave dust sublimation layer arising from an anisotropic radiation from the accretion disk \citep{KawaguchiEtAl_2010,KawaguchiEtAl_2011} or a turbulent structure near the nucleus arising from the interaction with other galaxies.

\section{CONCLUSION}\label{sec:con}

In this paper, we present a systematic spectral analysis of the CO ro-vibrational absorption band (4.67~\micron) toward ten nearby obscured AGNs observed with the \AKARI\ and \Spitzer\ space telescopes.
Using a gas model assuming LTE, slab geometry, and a single component gas, the CO column density and gas temperature were estimated for each target.
The average CO column density of the sample was found to be $\Nco\sim10^{19.5}~\psqcm$, which corresponds to a hydrogen column density of $\Nh\sim10^{23.5}~\psqcm$ if we assume a standard abundance ratio $[\mathrm{CO}]/[\mathrm{H}]\sim10^{-4}$.
This large column density indicates that the AGNs are heavily obscured.
On the other hand, the average temperature was found to be 360~K, which is much higher than the typical value in normal star-forming regions. 

The observed warm gas of a large column density cannot be represented by UV heating or shock heating.
The former can heat gas up to $10^3$~K, but its maximum heating depth is only $\Nco\sim10^{16}~\psqcm$, which is two orders of magnitude smaller than the observed values.
The latter can make gas warm ($\sim10^2$~K) up to $\Nco\sim10^{18}~\psqcm$, a column density comparable to the smallest \Nco\ we observed.
However, the low line-to-continuum ratios in the far-IR region of our sample galaxies indicate that, in addition to gas, dust is also heated up.
This does not occur under shock heating, in which gas and dust are thermally decoupled.
We conclude that the most convincing heating source is X-ray photons emitted from the nucleus.
This mechanism can heat gas up to even $\Nco\sim10^{20}~\psqcm$, which is large enough to account for the observed values over $10^2$~K.
This conclusion suggests that the region probed by the CO absorption should be in the vicinity of the nucleus.

A comparison with an X-ray spectral analysis \citep{BrightmanEtAl_2011a} shows that the hydrogen column density derived from conversion from the CO column density is 2--30 times smaller than that inferred from the X-ray analysis.
We conclude that the region probed by the near-IR CO absorption is located outside the X-ray emitting region.
Moreover, the close-to-unity covering factor of the CO gas suggested by the observed deep absorption indicates that the gas is close to the continuum source, which we hypothesize to be the dust sublimation layer at the inner rim of the obscuring material around the AGN.
In contrast to the comparison with the X-ray observations, the hydrogen column density derived from CO absorption is similar to or a bit larger than that calculated from the optical depth of the 9.7~\micron\ silicate dust absorption.
We conclude that CO-absorbing gas and silicate dust roughly coexist in the same region.

We reconfirmed a previously remarked-upon fact that not all Sy2 galaxies show CO absorption \citep{LutzEtAl_2004,GeballeEtAl_2009}.
The cryogenic phase \AKARI\ AGNUL results for the Sy2 galaxy IRAS 05189$-$2524 do not show any signature of the feature.
We speculate that the presence or absence of this absorption reflects the complex innermost geometry of the putative AGN torus, e.g., as a concave dust sublimation layer generated by anisotropic radiation from the accretion disk \citep{KawaguchiEtAl_2010,KawaguchiEtAl_2011} or a turbulent structure near the nucleus caused by interaction with other galaxies.

\acknowledgments

\ADD{The authors thank the anonymous referee for careful review and helpful suggestions.} 
The authors are grateful to H.~Spoon and J.~Cami for providing us with detailed information on their analysis of IRAS 00182$-$7112 together with insightful comments, and to F. Usui for supporting us in the data reduction of the \AKARI\ spectroscopy.
This work is supported by Grant-in-Aid for JSPS Research Fellows Grant Number JP17J01789 (S.B.) and for Scientific Research Grant Number JP26247030 (T.N.).

This research is based on observations with \AKARI, a JAXA project with the participation of ESA.
This work is based on observations made with the \textit{Spitzer Space Telescope}, obtained from the NASA/IPAC Infrared Science Archive, both of which are operated by the Jet Propulsion Laboratory, California Institute of Technology under a contract with the National Aeronautics and Space Administration.
This publication makes use of data products from the \textit{Wide-field Infrared Survey Explorer}, which is a joint project of the University of California, Los Angeles, and the Jet Propulsion Laboratory/California Institute of Technology, and \textit{NEOWISE}, which is a project of the Jet Propulsion Laboratory/California Institute of Technology.
\textit{WISE} and \textit{NEOWISE} are funded by the National Aeronautics and Space Administration.


\appendix

\section{Parameter dependence of the absorption profile}\label{sec:depen}

\begin{figure*}
\plotone{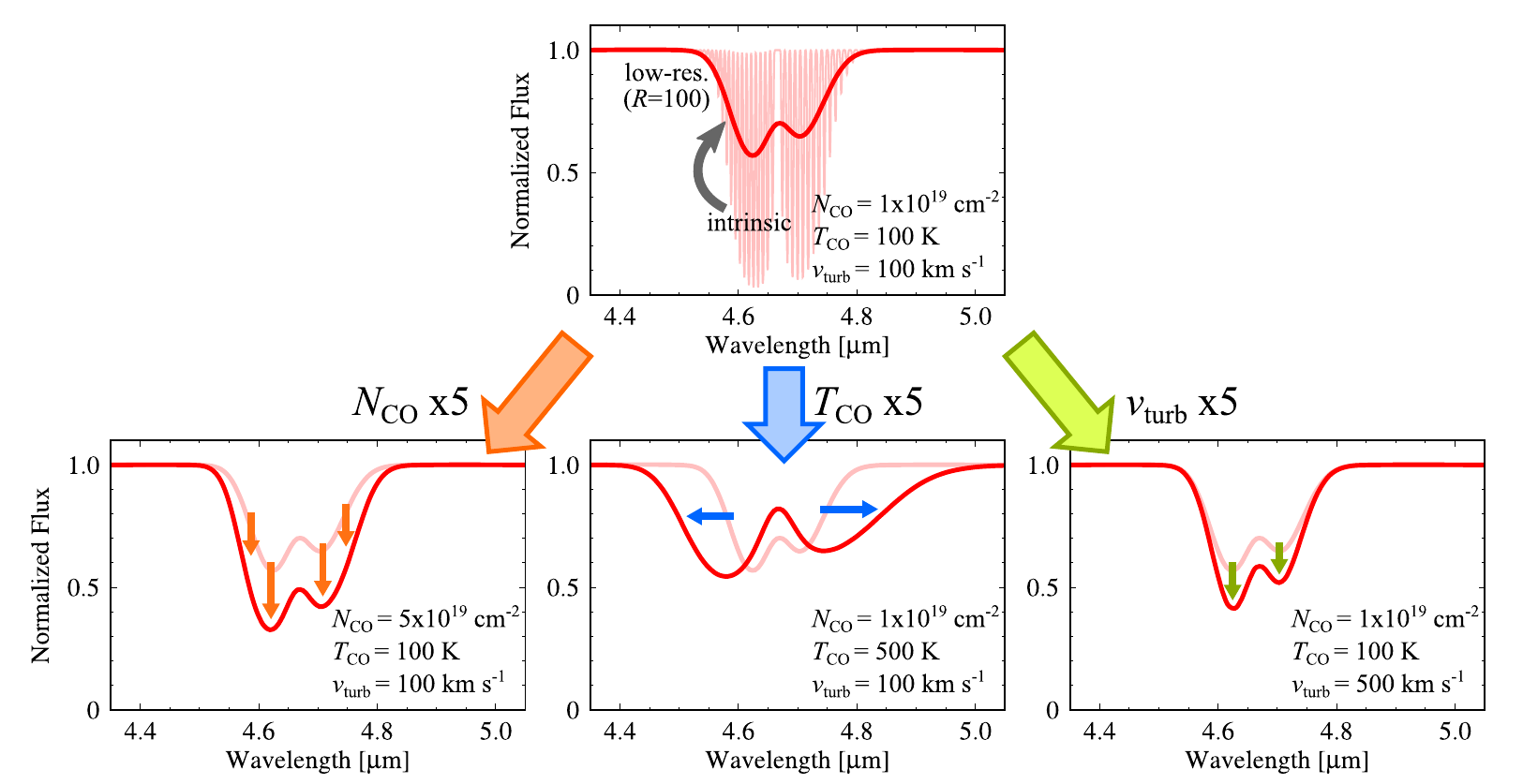}
\caption{Parameter dependence of the CO absorption model. The top panel compares the intrinsic model spectrum for the parameters noted at the right bottom corner with a blurred spectrum for a low spectral resolution ($R=100$). The bottom panels illustrate the changes in the absorption profile from the top panel when one parameter is multiplied by a factor of five.\label{fig:param_depen}}
\end{figure*}

Figure \ref{fig:param_depen} describes how our model depends on three parameters.
If the column density, \Nco, increases, the absorption profile becomes deeper because the optical depth is proportional to \Nco.
If the temperature, \Tco, increases, the profile becomes wider because higher rotational levels are populated.
If the velocity width, \vturb, increases, the profile becomes deeper at the peaks of the $P$- and $R$-branches because the equivalent width is broadened only when the absorption is saturated.
When the absorption is weak, we cannot determine \vturb\ from the model fitting unless we resolve rotational levels with a high spectral resolution.
In this paper, however, we do not face this problem in most of the targets because they show sufficiently deep absorption profiles.
We restrict the range of \vturb\ to less than 300~\kmps\ because values higher than this upper limit do not substantially affect the profile as they involve two neighboring rotational lines to completely blend with each other.

\section{$\Delta\chi^2$ map}\label{sec:Dchi2}

Figure \ref{fig:chi2map} shows a set of color maps of the $\Delta\chi^2$ value for all of the targets.
The maps clearly show the degeneracy among the parameters.
Large values of \Nco\ anti-correlates with \Tco\ and \vturb.
The \Nco\ versus \Tco\ anti-correlation can be interpreted as follows.
At large \Nco, the absorption is nearly saturated.
Because the depths at the two peaks of the absorption profile do not change significantly with \Nco, the depths at the band wings become more important in the determination of the solution.
However, high values of \Tco\ instead of \Nco\ can also deepen the band wings, which results in the anti-correlation between \Nco\ and \Tco.
On the other hand, the \Nco\ versus \vturb\ anti-correlation originates because the two parameters similarly deepen the band profile when the absorption is saturated.

\begin{figure*}
\plotone{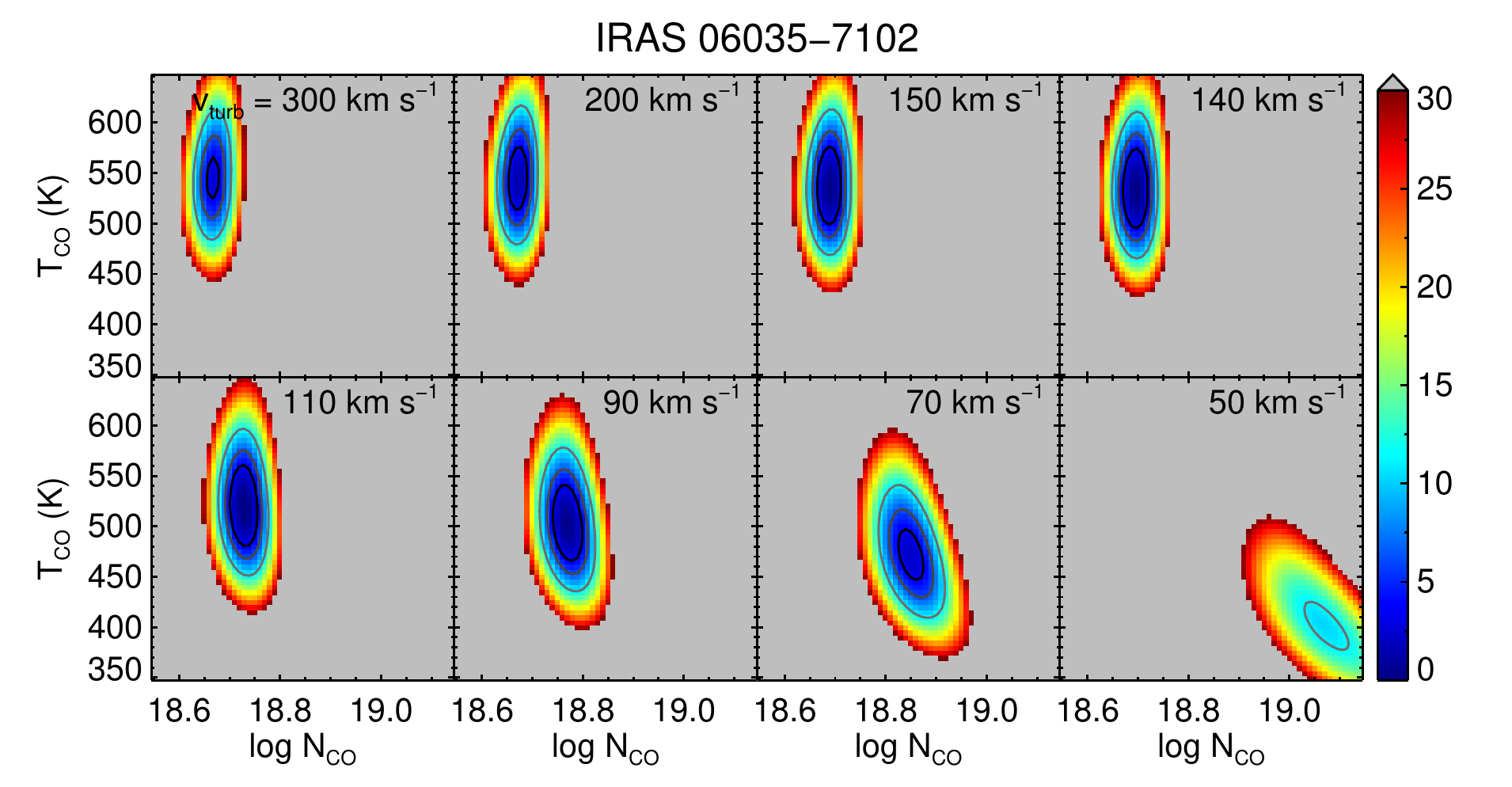} 
\plotone{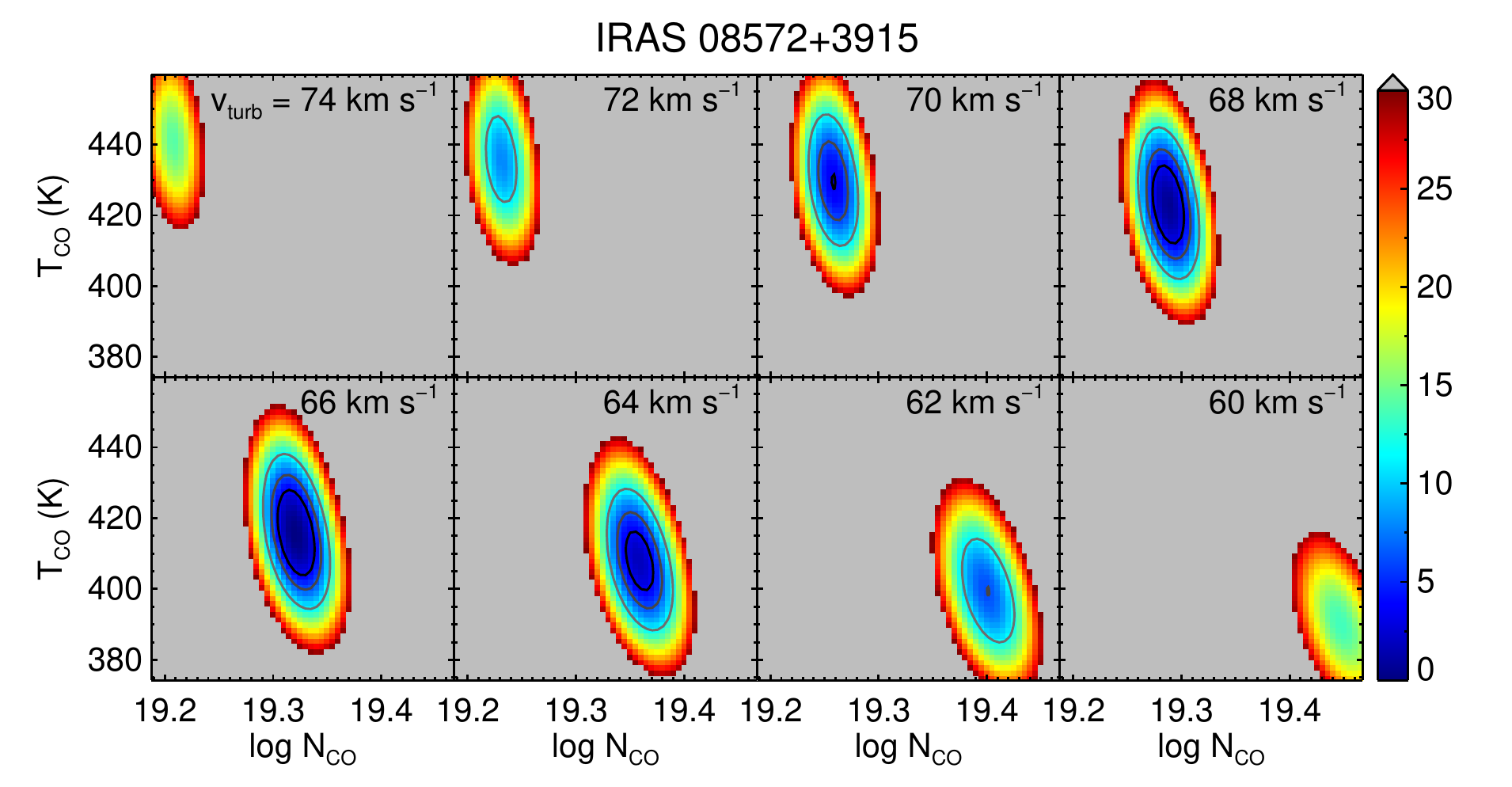} 
\caption{Color maps of the $\Delta\chi^2$ values of the best-fit models shown in Figure \ref{fig:bestfit} whose parameters are tabulated in Table \ref{tab:bestfit}. Each panel shows a slice of the three-dimensional parameter space at the turbulent velocity \vturb\ noted at the top right corner. From inside to outside, the closed solid curves denote the 68\%, 90\%, and 99\% joint confidence levels, respectively.\label{fig:chi2map}}
\end{figure*}
\addtocounter{figure}{-1}
\begin{figure*}
\plotone{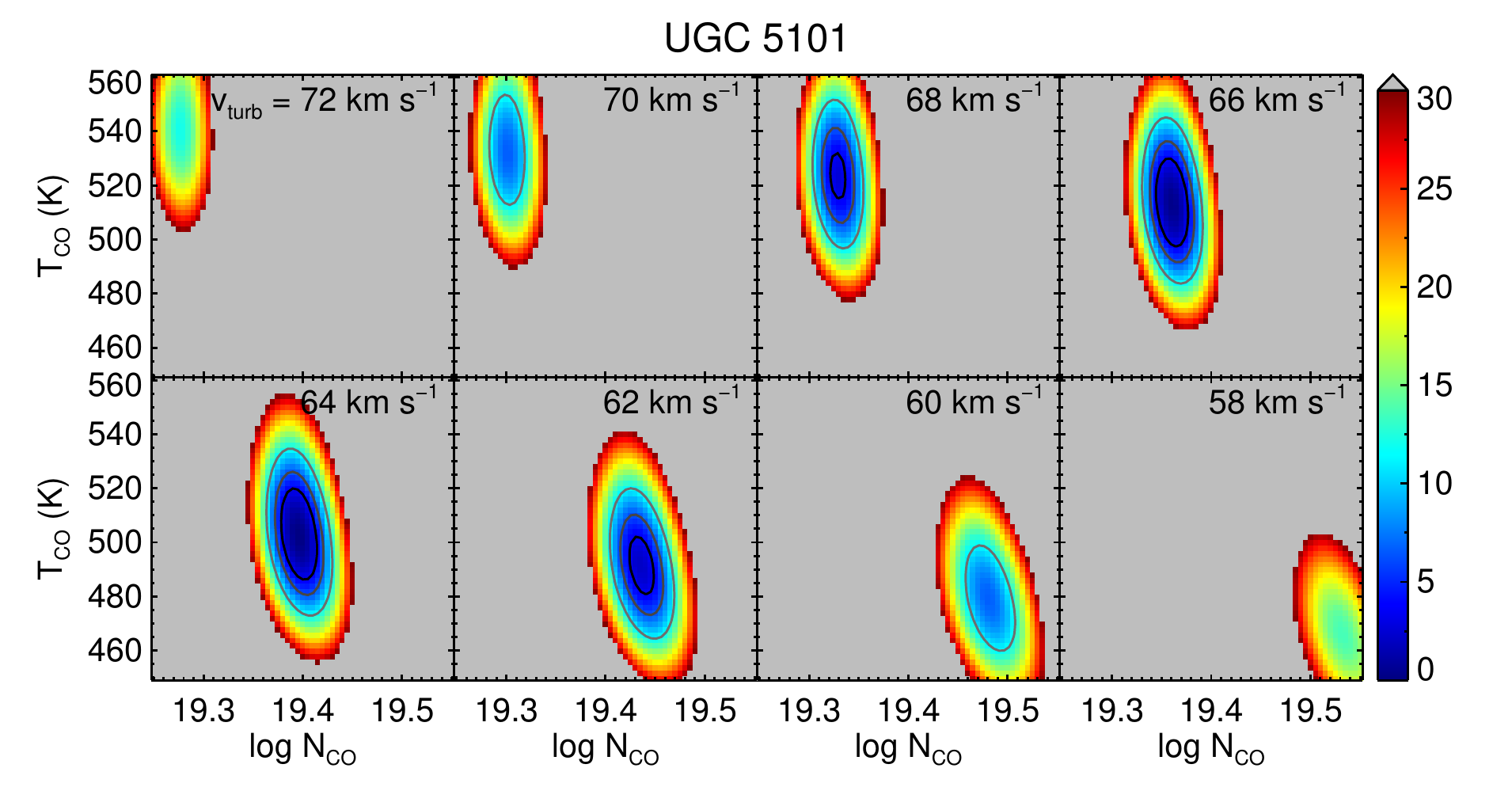} 
\plotone{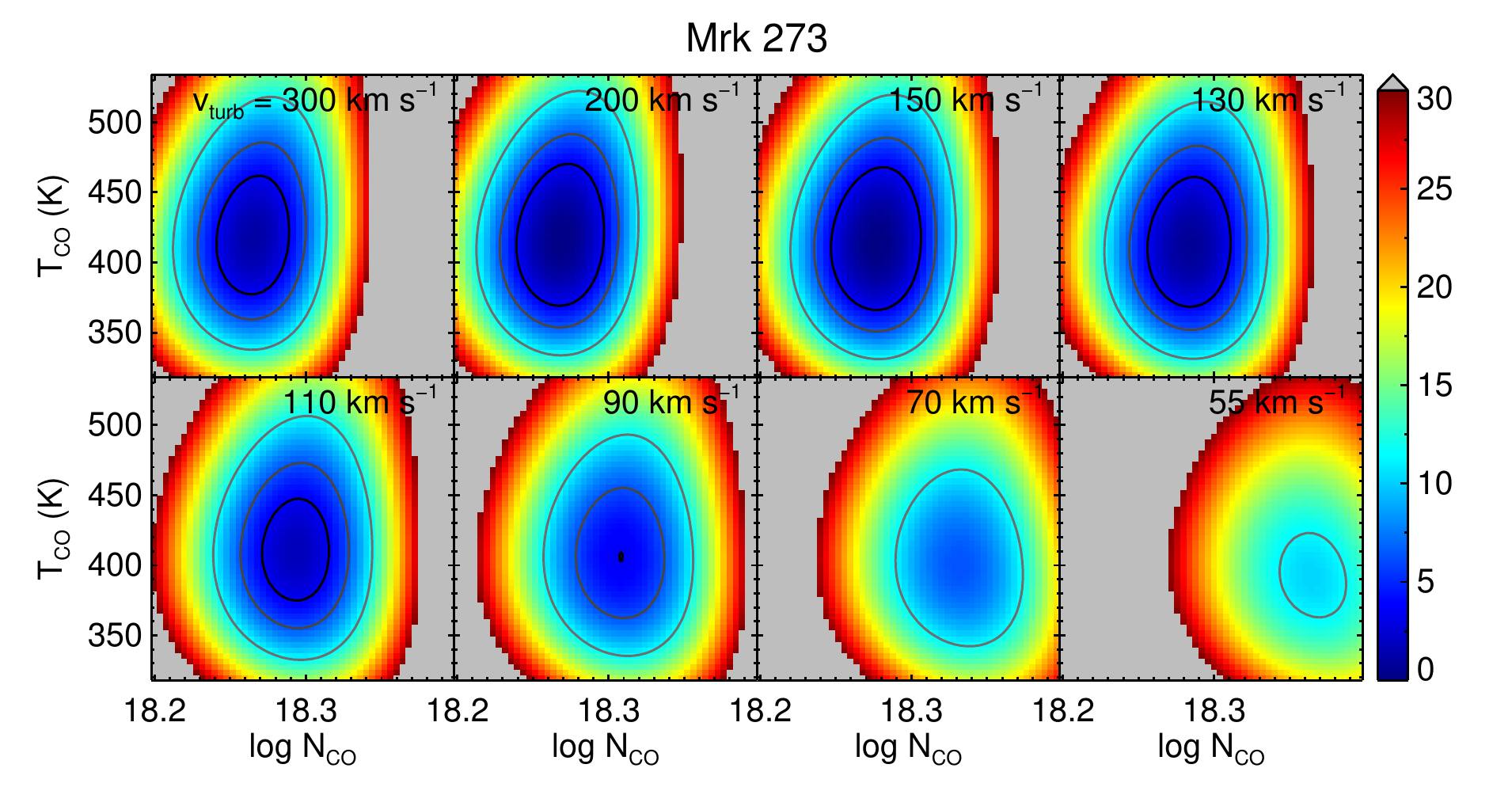} 
\caption{\textit{(Continued)}}
\end{figure*}
\addtocounter{figure}{-1}
\begin{figure*}
\plotone{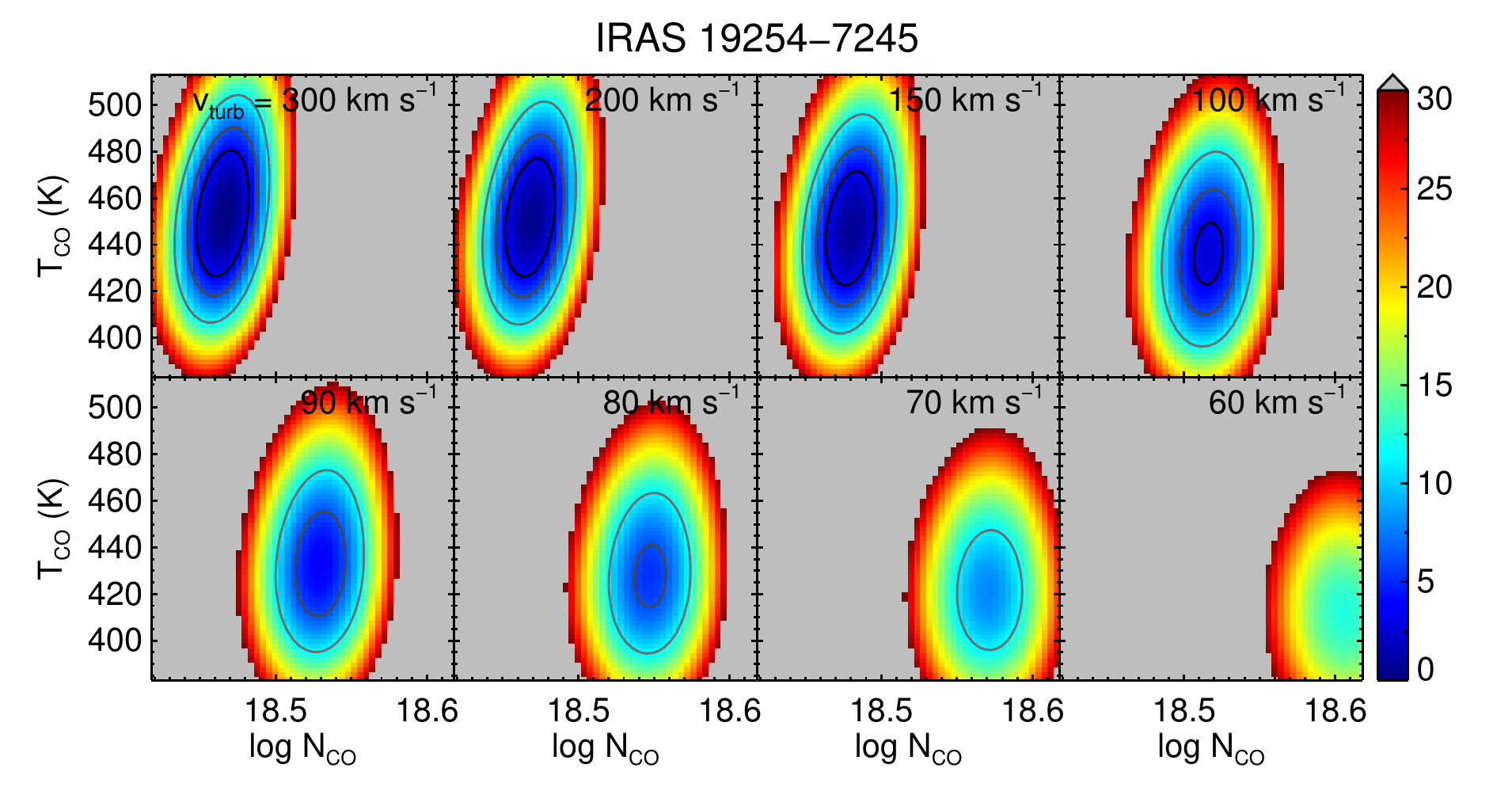} 
\plotone{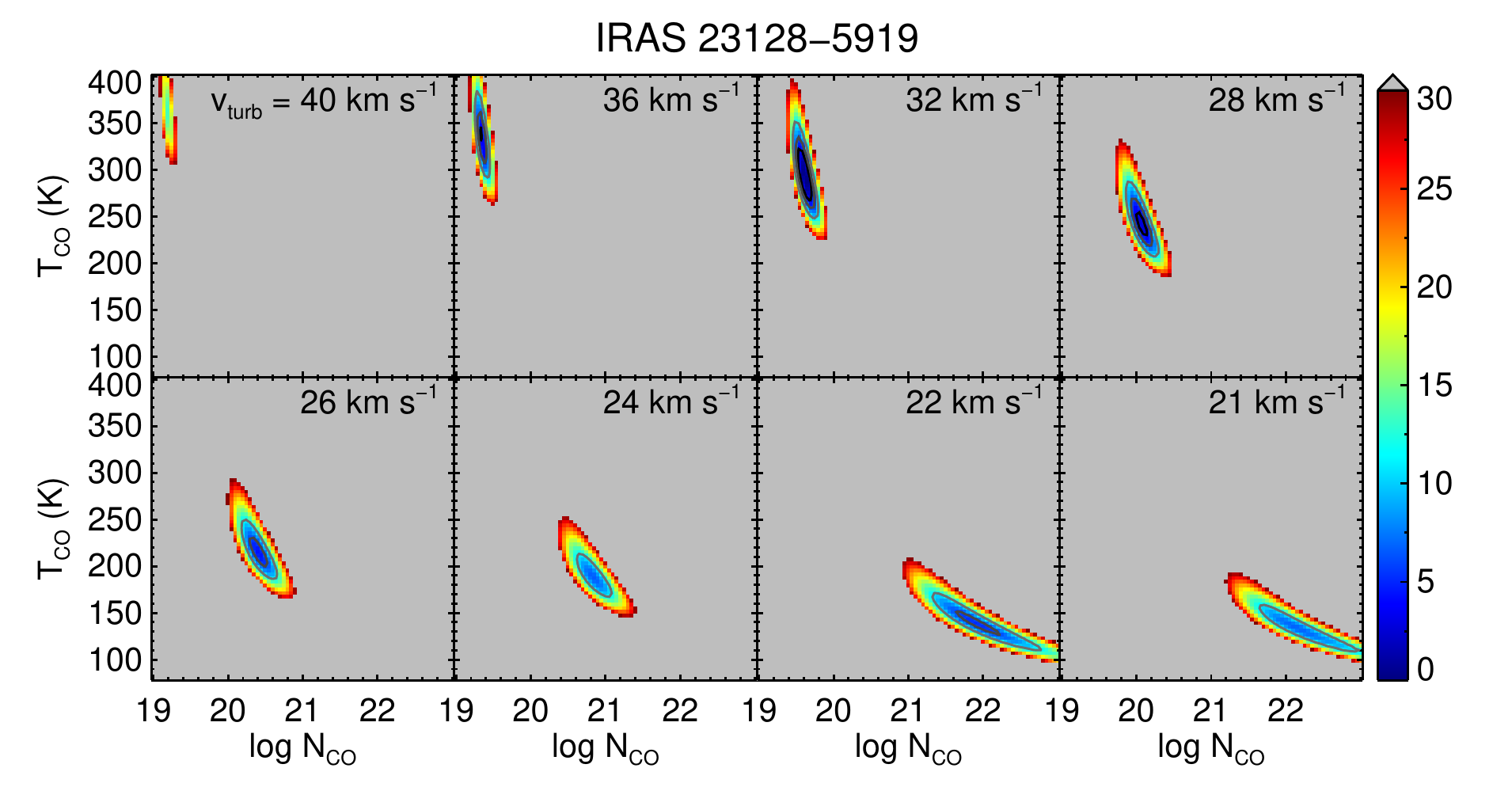} 
\caption{\textit{(Continued)}}
\end{figure*}
\addtocounter{figure}{-1}
\begin{figure*}
\plotone{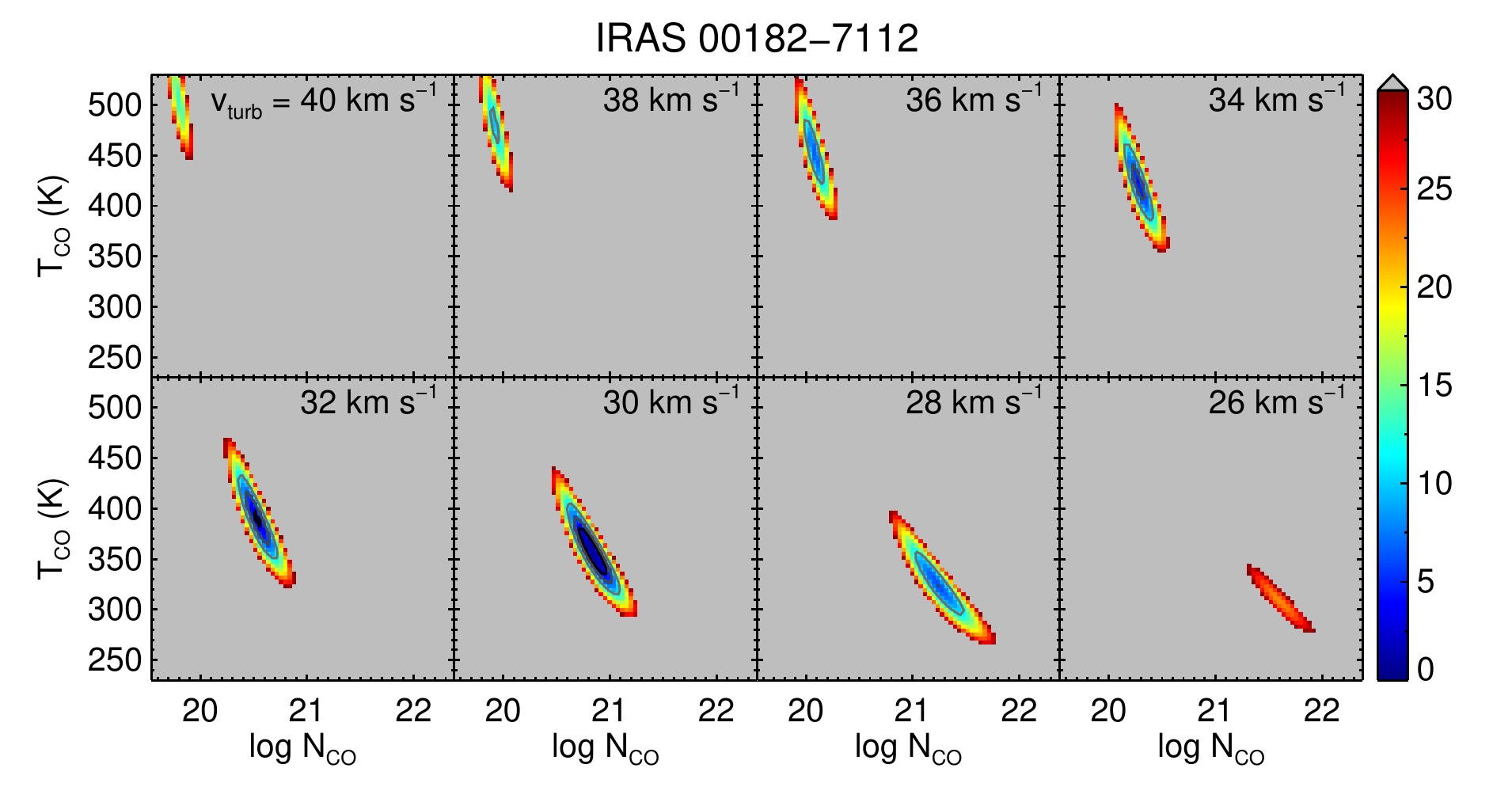} 
\plotone{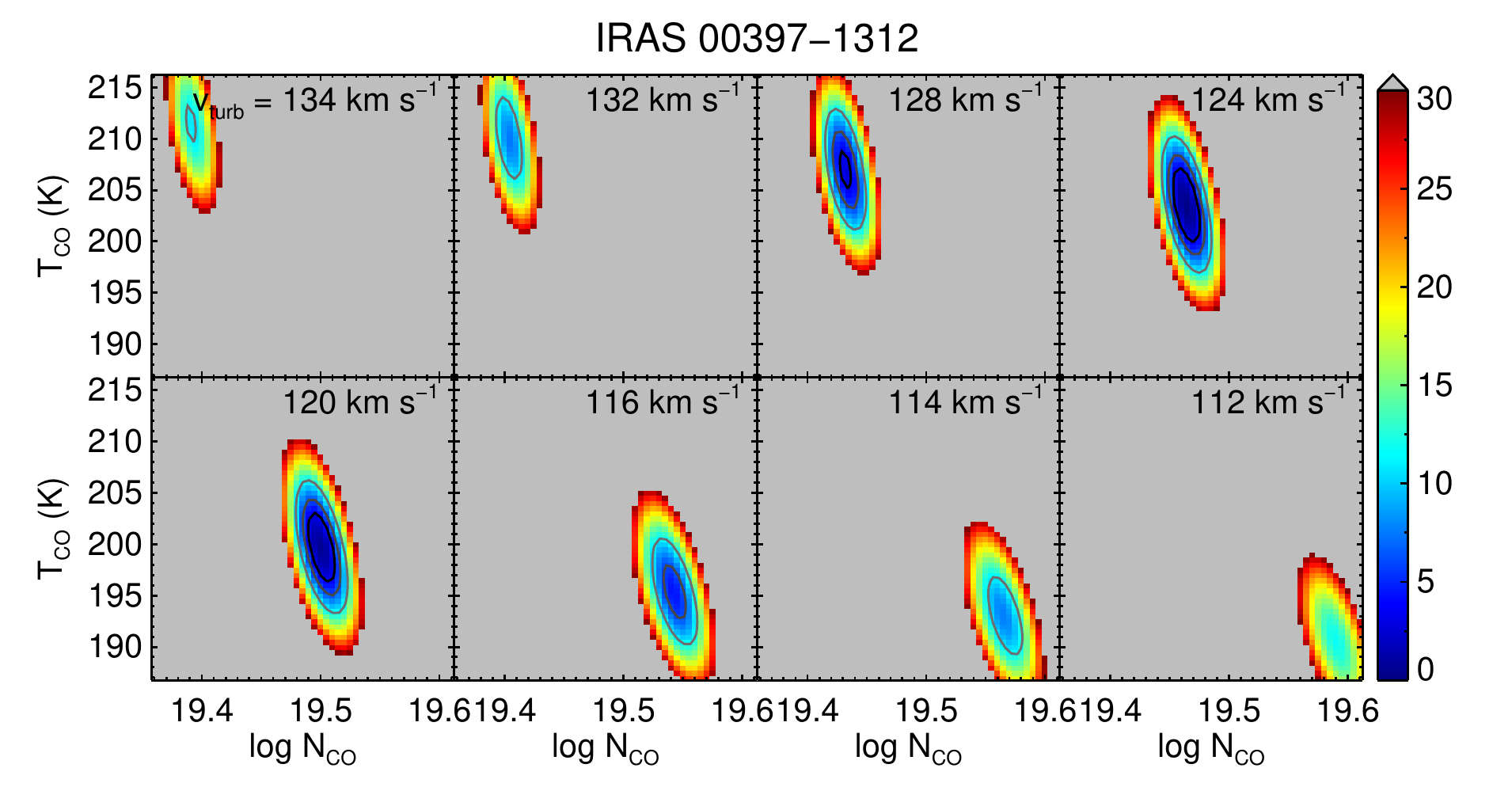} 
\caption{\textit{(Continued)}}
\end{figure*}
\addtocounter{figure}{-1}
\begin{figure*}
\plotone{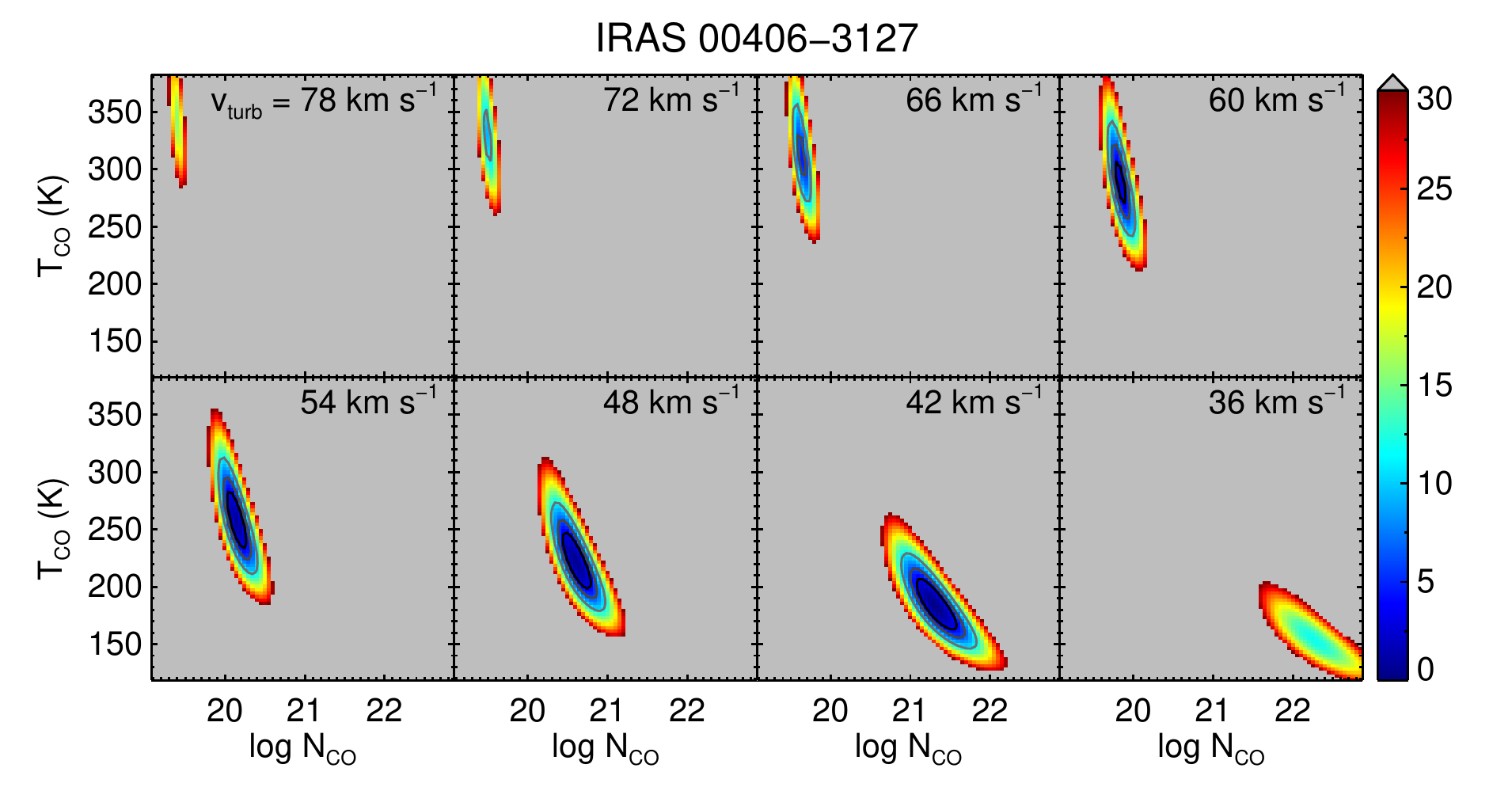} 
\plotone{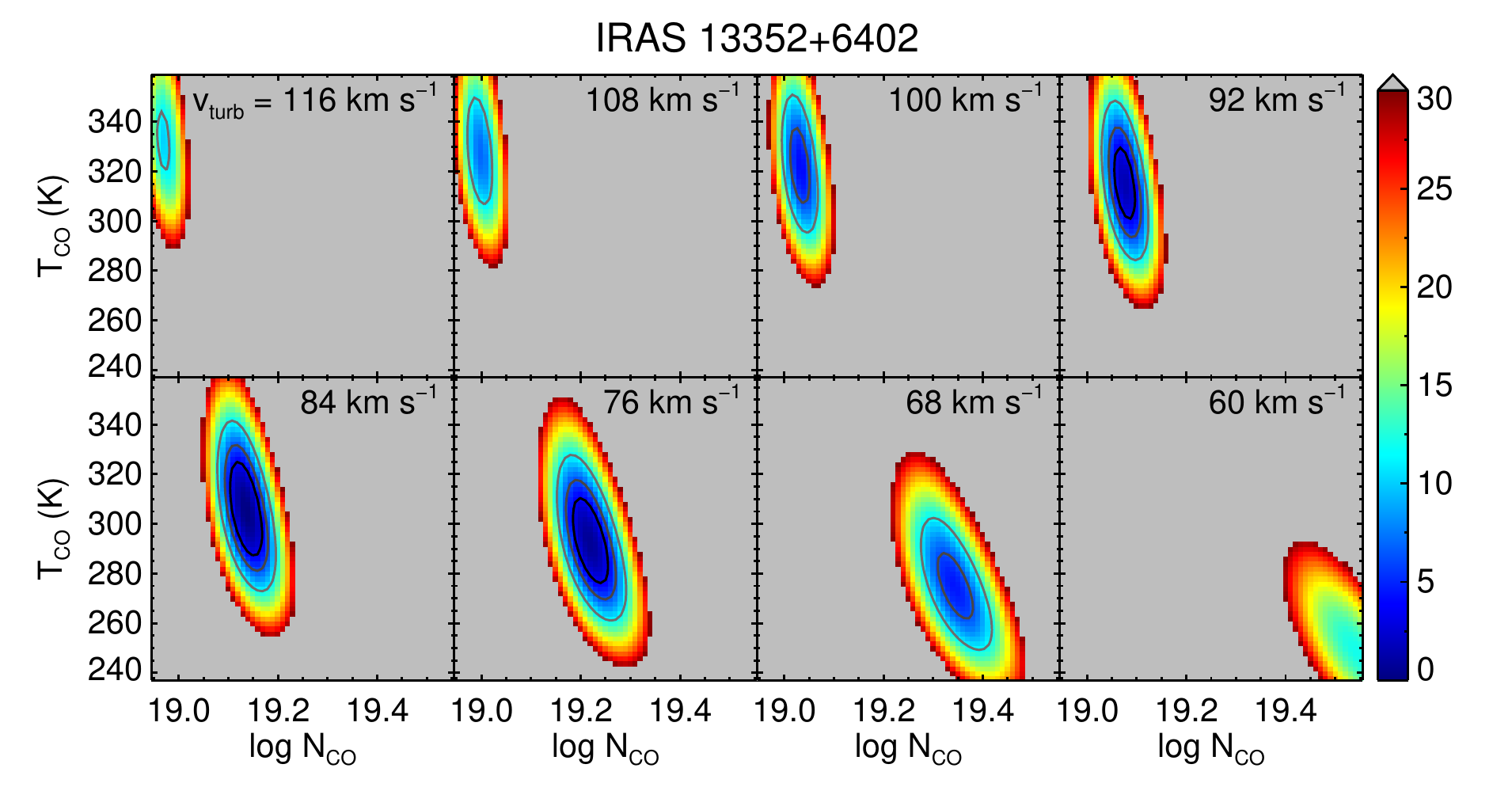} 
\caption{\textit{(Continued)}}
\end{figure*}

\section{Comparisons with Previous CO Analyses}\label{sec:previous}

\subsection{IRAS 00182$-$7112}

Our best-fit model for IRAS 00182$-$7112 (or IRAS F00183$-$7111) had a large column density and did not coincide with that of \citetalias{SpoonEtAl_2004b}, who had performed a similar gas model fitting for a \Spitzer/IRS spectrum based on an earlier IRS calibration.
Our best-fitting parameters were, as tabulated in Table 3, $\Nco=10^{21.2}~\psqcm$, $\Tco=329~\mathrm{K}$, and $\vturb=28~\kmps$, whereas those found by \citetalias{SpoonEtAl_2004b} were $10^{19.5}~\psqcm$, 720~K, and 50~\kmps.
The 99\% joint confidence region of our best-fit did not include the solution of \citetalias{SpoonEtAl_2004b}.
However, in the analysis of \citetalias{SpoonEtAl_2004b}, another solution with $\Nco=10^{21.5}~\psqcm$, $\Tco=400~\mathrm{K}$, and $\vturb=25~\kmps$, which is similar to our result, was also found (H. Spoon \& J. Cami, private communication).
We therefore finally adopted our obtained large-column solution even though this solution has an uncertainty in the amount of gas hidden behind the $\tau=1$ surface.

\subsection{IRAS 08572$+$3915}

IRAS 08572$+$3915 comprises two nuclei, with the one in the northwest (NW) being brighter than the one in southeast (SE) by 2.2 magnitudes at 2.2~\micron\ \citep{ScovilleEtAl_2000a}.
The CO ro-vibrational absorption lines in IRAS 08572$+$3915 NW were first detected by \citet{GeballeEtAl_2006} using UKIRT, and then observed with higher quality by \citetalias{ShirahataEtAl_2013} using Subaru.
\citetalias{ShirahataEtAl_2013} obtained a high spectral resolution spectrum ($R\sim5000$, $\Delta v\sim60~\kmps$) with individual absorption lines resolved and revealed a velocity profile comprising three distinct components centered at 0, $-$160, and $+$100~\kmps\ relative to the systemic velocity of the galaxy.
The authors analyzed the spectrum under the assumption of a three-component gas.
Note that the numbers below are quoted from the full coverage case of \citetalias{ShirahataEtAl_2013}, which is equivalent to our  assumption, although the authors also discussed a partial coverage case.
The first component was the blueshifted component, which had the largest CO column density of $2.7\times10^{18}~\psqcm$, a warm temperature of 325~K, and a velocity width broader than 200~\kmps.
The second component was at the systemic velocity, with a column density of $5.7\times10^{17}~\psqcm$ and a temperature of 23~K.
The third was the redshifted component, which was weaker than the other two components ($\sim10^{17}~\psqcm$) and had the highest temperature ($\sim700~\mathrm{K}$).

Although the \AKARI\ observation used in this work did not spatially resolve the two nuclei, we assume that the contribution from the SE nucleus at around 4.67~\micron\ is negligible as well as that at 2.2~\micron, as SE is also much fainter in mid-IR wavelengths \citep{SoiferEtAl_2000}.
Our best-fitting column density, $\Nco=10^{19.3}~\psqcm$, and temperature, $\Tco=417~\mathrm{K}$, were $\sim7$ times larger and $\sim1.3$ times higher, respectively, than those of the most prominent component of \citetalias{ShirahataEtAl_2013}.
Below, we discuss the source of this inconsistency.

The \AKARI\ spectrum used in this study and the Subaru spectrum used in \citetalias{ShirahataEtAl_2013} have quite different wavelength coverages.
The former covered only the $R$-branch, while the latter, which was limited by the atmospheric window, covered an almost opposite wavelength range corresponding to the $P$-branch.
Thus, the discrepancy in the column densities found from the $R$- and $P$-branches can be attributed to the absorption profile becoming ``asymmetric'' and deviating from the current model prediction.
Although it is possible that the effects mentioned in Section \ref{sec:beyond} can cause such an asymmetric absorption profile, these effects are weak to explain the difference.

We propose that the most plausible interpretation for the discrepancy is that \citetalias{ShirahataEtAl_2013} underestimated the continuum level in the calculation of the equivalent widths.
As the Subaru spectrum used by \citetalias{ShirahataEtAl_2013} did not cover featureless wavelength regions, the authors determined the continuum level under the assumption that high-intensity peaks correspond to zero-absorption intensities.
However, this determination is not trivial.
The absorption depth in the Subaru spectrum coincides with that observed with \AKARI\ if the actual continuum level is higher than that adopted by \citetalias{ShirahataEtAl_2013} by 30\%.
In this case, the equivalent width of each rotational line increases, and the excitation temperature of the dominant component obtained from the population diagram becomes 480~K.
This change follows the trend in the discrepancy between our analyses and those of \citetalias{ShirahataEtAl_2013}.
To confirm this explanation, we require a seamless spectrum that continuously covers both the $P$- and $R$-branches.
Such a spectrum can be acquired using the upcoming \textit{James Webb Space Telescope}.

\end{document}